\documentclass[review]{elsarticle}

\usepackage{lineno,hyperref}
\usepackage{algorithm}
\usepackage{algorithmic}
\usepackage{IEEEtrantools}
\usepackage[cmex10]{amsmath}
\usepackage{mathtools}
\usepackage{xcolor}
\usepackage{soul}
\DeclareMathOperator*{\argmin}{argmin}
\if CLASSOPTIONcompsoc
  \usepackage[caption=false,font=normalsize,labelfont=sf,textfont=sf]{subfig}
\else
  \usepackage[caption=false,font=footnotesize]{subfig}
\fi

\usepackage{amsfonts}
\usepackage{graphicx}

\modulolinenumbers[5]

\journal{Journal of Pervasive and Mobile Computing}

\begin{document}

\begin{frontmatter}

\title{Signal Strength based Scheme for Following Mobile IoT Devices in Dynamic Environments}
%\tnotetext[mytitlenote]{Fully documented templates are available in the elsarticle package on \href{http://www.ctan.org/tex-archive/macros/latex/contrib/elsarticle}{CTAN}.}

%% Group authors per affiliation:
%\author{Thomas Lagkas\fnref{myfootnote}}
%\address{Radarweg 29, Amsterdam}
%\fntext[myfootnote]{Since 1880.}

%% or include affiliations in footnotes:
%\author[mymainaddress,mysecondaryaddress]{Elsevier Inc}
%\ead[url]{www.elsevier.com}

%\author[mysecondaryaddress]{Global Customer Service\corref{mycorrespondingauthor}}
%\cortext[mycorrespondingauthor]{Corresponding author}
%\ead{support@elsevier.com}

%\address[mymainaddress]{1600 John F Kennedy Boulevard, Philadelphia}
%\address[mysecondaryaddress]{360 Park Avenue South, New York}

\author[add1,add2]{Thomas Lagkas}
\ead{t.Lagkas@sheffield.ac.uk}
\author[add1]{George Eleftherakis}
\ead{g.eleftherakis@sheffield.ac.uk}
\author[add1]{Konstantinos Dimopoulos}
\ead{k.dimopoulous@sheffield.ac.uk}
\author[add3]{Jie Zhang}
\ead{jie.zhang@sheffield.ac.uk}

\address[add1]{Computer Science Department, The University of Sheffield International Faculty - CITY
College, Thessaloniki, Greece}
\address[add2]{Department of Computer Science, International Hellenic University, Kavala Campus, Greece}
\address[add3]{Department of Electronic and Electrical Engineering, The University of Sheffield, Sheffield, UK}

\begin{abstract}
The increased maturity level of technological achievements towards the realization of the Internet of Things (IoT) vision allowed sophisticated solutions to emerge, offering reliable monitoring in highly dynamic environments that lack well-defined and well-designed infrastructures.
%, such as in the case of disaster scenarios.
In this paper, we use a bio-inspired IoT architecture, which allows flexible creation and discovery of sensor-based services offering self-organization and self-optimization properties to the dynamic network, in order to make the required monitoring information available. The main contribution of the paper is the introduction of a new algorithm for following mobile monitored targets/individuals in the context of an IoT system, especially a dynamic one as the aforementioned. The devised technique, called Hot-Cold, is able to ensure proximity maintenance by the tracking robotic device solely based on the strength of the RF signal broadcasted by the target to communicate its sensors' data. Complete geometrical, numerical, simulation, and convergence analyses of the proposed technique are thoroughly presented, along with a detailed simulation-based evaluation that reveals the higher following accuracy of Hot-Cold compared to the popular concept of trilateration-based tracking. Finally, a prototype of the full architecture was implemented not only to demonstrate the applicability of the presented approach for monitoring in dynamic environments, but also the operability of the introduced tracking technique.
\end{abstract}


\end{frontmatter}

%\linenumbers

\section{Introduction}

It is a fact that the world frequently witnesses emergent situations which are related with major problems in infrastructures due to natural causes or the result of intentional or accidental human actions.
%It is a fact that the last two decades were marked with world-wide emergency disasters, either natural disasters or the result of intentional or accidental human actions. 
Apart from their tremendous impact they all shared another characteristic: they reduced communication among humans. Modern means of communication were significantly degraded which rendered any rescue operations significantly harder.
One of the main challenges in emergency management is to perform efficient monitoring and coordination. This aspect heavily relies on the communication between involved actors and the availability of information by monitoring victims' vital signs and also other environmental measurements. Traditional approaches in emergency scenarios lean on some sort of centralized or well and in advance engineered solution. However, as recent emergencies demonstrated, there is a need for communication mechanisms which do not rely on centralized systems and infrastructure. Such alternative solutions can be used as a fall-back mechanism in the case that primary systems fail.

At the same time, a remarkable maturity in recent technological advancements have led to the Internet of Things (IoT) as the most promising achievement towards smart solutions in a variety of applications. In this context, there is an inevitable need for scalable IoT architectures that offer advanced positioning, localization and context awareness based services for sophisticated applications enabling smart solutions (e.g., e-health, smart cities, smart emergency management, etc.). Such solutions need to be available even in extremely dynamic and GPS-denied environments, allowing the deployment of flexible and autonomous sensor networks, especially in situations where well-defined infrastructures do not exist or are not preferable.

In this sense and in order to enable effective management and communication, there is a need for monitoring security and rescue forces personnel, victims, and other actors offering useful information in environments with no-well designed or crippled infrastructures. 
A common problem in such scenarios is also tracking down continuously a target that transmits useful information aiming to follow it maintaining proximity. At the same time, another critical point is the need of using resource limited components in such an attempt and preserve energy. 

This work has been developed as part of an IoT architecture introduced in \cite{eleftherakis_architecting_2015} that is capable of providing autonomous sensor-based distributed services. This architecture is based on bio-inspired principles found in natural systems to achieve the required robust behavior. In more detail, it is based on networked autonomous software agents which are serving information collected by interconnected IoT devices in weakly structured environments. The agents’ behaviour, such as service discovery and self-organization, is adapted to the patterns of service consumers’ (i.e. users) requests. The considered system is described in Section 3. However, it is noted that the scheme presented in this paper is architecturally agnostic, which means that it can be applied either to the aforementioned IoT architecture, or to any other IoT system, or as a standalone application.

The main contribution of this paper is the introduction of a new scheme that allows a robotic device to efficiently follow a monitored target solely based on the RF signal the latter broadcasts for communicating its sensors' data. The respective algorithm is called "Hot-Cold" and is thoroughly described, analyzed, and evaluated. A distinctive characteristic of the proposed solution which enhances robustness is the ability to maintain proximity without the necessity of identifying target's position. This tracking scheme is realized as a significant component of an original IoT system, which enables agile services creation and discovery for the provision of flexible access to real-time monitoring information in dynamic environments. System feasibility and effectiveness is ensured through the implementation and testing of a complete prototype. The conducted evaluation shows the efficiency of the proposed technique in following a radio-emitting IoT device solely based on the strength of its communication signal, outperforming the well-known trilateration-based tracking in realistic shadowing conditions.

The rest of the paper is structured as follows. The next section presents background approaches in the field of location based services in IoT, emphasizing on related work on target tracking. Section III presents the adopted IoT system architecture, describing all main components. The proposed Hot-Cold algorithm is detailed and analyzed in Section IV. The following section documents the evaluation of the tracking scheme, discusses simulation results, and presents the system prototype. Finally, Section VI concludes the paper and provides insights for future work.

\section{Background and Related Work}

In this section we review the main concepts of this work and we focus on target tracking approaches.

\subsection{Internet of Things and Location-based Services}
Internet of Things (IoT) is a global collection of physical and virtual devices, and all related infrastructure, exchanging information and providing services to each other and to the people who use it \cite{ray_survey_2016}. Many services provided are Location Based Services (LBS). Depending on the prospective one can see it, LBS involves the use of location information of the target or the provider of the service \cite{kupper_location-based_2005}. Location Service (LS) is the process of defining the location of an asset, and is a crucial part of any LBS, and is usually done by GPS or other sensor embedded in the device. These sensors can provide a very accurate location, but are costly in power consumption. This can become an issue when non-accurate location information will suffice, but there are power consumption limitations. With each device added to the IoT, global processing power is increased linearly, but communication channels between devices is following a much higher progression. Whatever the case, the communication technology is some form of wireless communication \cite{ray_survey_2016}. This means that information transmission between the involved devices takes place by default anyway, hence, the transmission itself can be exploited to provide an inexpensive means of LS.

\subsection{Localization Techniques Overview} 
Localization is a topic that has predominated mobile robotics for a very long time. According to \cite{siegwart_introduction_2004}, localization is the problem of defining the spacial information (location, velocity and orientation) of a mobile robot in space. 

By  far the simplest technique is that of odometry. This involves measuring the movement of the robot (using rotational sensors on the wheels, or inertial sensors) to determine the change in position on regular intervals. However, as this approach is open-loop, it requires validation regularly as errors in measurements soon accumulate. 

Another technique is based on using beacons. Beacon systems can be used to actively or passively determine the location of the robot, through triangulation or trilateration. The active or passive component is determined by whether the transmitting beacons are located on fixed known locations with the receiver on the robot or vice versa. Triangulation uses the estimated angles between the robot and the beacons to calculate accurately the location of the robot. Trilateration is using the estimated distances from the robot to the beacons to achieve the same goal. In both cases, complex mathematical equations have to be solved, a process that can become computationally heavy especially if repeated often.

Other, more advanced localization techniques require visual recognition of artificial landmarks, or visual tracking of the robot itself through cameras on the ceiling. Both techniques require image processing, thus vision equipment and visual line of sight, and therefore are outside the scope of this paper. For a full review of these techniques and more, see \cite{hightower_survey_2001}. 

Finally, WiFi Positioning Systems (WPS) use geo-references radio maps of areas, to provide positioning information \cite{zhuang_autonomous_2014}. The accuracy of this approach, however, has a heavy cost as it is limited within a predefined area, and the process of building in advance the radio map is required \cite{zhuang_evaluation_2016}.

\subsection{Tracking Strategies Overview} 
Tracking involves knowing the location of a target, and navigating towards it. This could involve the avoidance of obstacles or path making. Control laws for tracking a target can be formulated with estimations of distance from the target \cite{fidan_adaptive_2013, bopardikar_pursuit_2008, namaki-shoushtari_switched_2011, chaudhary_capturing_2013, cao_circumnavigation_2013, teimoori_equiangular_2010}. In \cite{fidan_adaptive_2013}, a tracking strategy is presented, where the distance from the target is estimated over time, by using the strength of a received signal. Consecutive distance estimations are then used to define a control law that defines the motion of the tracking agent to follow the target object. Another strategy that uses distance measurements between target and tracker can be based on the trilateration method, we discussed at the previous subsection. According to \cite{bopardikar_pursuit_2008}, by using this strategy it is possible for the tracker to locate a moving target in a bounded time, given that the target's speed is up to half the speed of the tracker. A similar approach is followed in \cite{namaki-shoushtari_switched_2011}, but there the problem is solved using orientation and distance information from the target, while in \cite{cao_circumnavigation_2013} and \cite{teimoori_equiangular_2010}, the problem is solved without knowledge of the orientation of the target, but only the estimated distance from it and its derivative.

\subsection{Target Tracking Techniques for Dynamic Networks}
Recently, tracking of mobile devices in dynamic networks has attracted a lot of interest, due to the promising applications in various use cases. Such dynamic networks include indoor tracking scenarios as well as outdoor tracking in the context of a wireless sensor network deployment. A number of related algorithms and solutions have been introduced, exploiting the properties of the emitted electromagnetic signal. 

A well-known tracking technique for such environments of dynamic signal variations is based on Linear Least Squares (LLS). LLS is frequently adopted in localization scenarios for the estimation of parameters which are initially unknown, by adjusting the observed parameters. The specific technique is characterized as linear or non-linear depending on the type of the derived system of equations. In more detail, in a 2-dimension setup, the target location can be estimated employing at least three lateral or angular measurements typically taken by reference points or nodes of known coordinates within the network. Authors in \cite{gezici_performance_2008} have applied linear LLS to enhance trilateration. In such a case, the location of the target is derived via the following formula:
\begin{equation}
 \hat{\pmb{x}} = (\pmb{A}^{T}\pmb{A})^{-1}\pmb{A}^{T}\hspace{1mm}\pmb{b}
 \label{eq:LLS}
\end{equation}
where $\hat{\pmb{x}}$ is the vector of the target coordinates' estimated corrections, $\pmb{A}$ is the \emph{design matrix}, and $\pmb{b}$ is the vector with the \emph{residual observations}. However, in a highly dynamic environment where the target and the tracking device have high relative speed, it becomes very challenging for LLS to make frequent and accurate estimations, since too frequent observations tend to be highly correlated resulting in a singular $\pmb{A}^{T}\pmb{A}$.

Another approach for tracking in dynamic environments, where deterministic modelling is very difficult, is the Particle Filter Localization (PFL) technique \cite{huang_analytically-guided-sampling_2013}. According to the respective algorithm, a number of random samples (particles) are initially generated. The target's state, as well as the particles' state, is defined by a set of parameters, such as position and velocity. In an iterative manner, observations are periodically collected and the particles' states are accordingly updated in an effort to estimate target's actual motion. Each particle is associated with an adjustable normally distributed weight, which indicates the probability to match the target's actual state, resulting in particles which converge to that target's state. In summary, PFL is a probabilistic algorithm with promising performance when a high number of particles are considered, which comes with the cost of increased computational requirements. 

The use of Kalman Filtering (KF) for target tracking within sensor networks has risen as an attractive technique adopted by a number of related algorithms with different variations, such as Extended Kalman Filtering (EKF) and Distributed Kalman Filtering (DKF). A promising approach is introduced in \cite{olfati-saber_distributed_2008}, where a message-passing version of the Kalman-Consensus Filtering (KCF) is proposed to facilitate distributed tracking of a maneuvering target in a network of sensors with limited range. The authors introduce a hierarchical architecture to collect and distribute the estimates of the micro Kalman filters in a Peer-to-Peer (P2P) sensor network. The microfilters update the states based on the received feedback and fuse their outputs as messages to other peers. The resulted P2P/Hierarchical architecture is shown to achieve high tracking performance, however, no sensors' mobility is considered.

A very interesting approach in dynamic target tracking, which is closely related with the scenarios considered in our work, is flocking control in mobile sensor networks. In such dynamic networks, nodes are typically mobile robotic devices equipped with various sensors. An adaptive flocking control algorithm is introduced in \cite{la_dynamic_2012}, where a group of mobile sensors cooperate and adjust connectivity and topology formation to the current network environment. Moreover, a multiple dynamic target tracking algorithm, called Seed Growing Graph Partition (SGGP), is proposed to address the merging/splitting problem. Both presented algorithms rely on graph network modelling and forces which either attract or repel the nodes. The conducted experimental tests verify the effectiveness of the algorithms, however, the focus is on the group adaptation rather than the explicit target tracking process.

Lastly, a promising target tracking solution for short range dynamic networks is based on Ultra-Wide Band (UWB) signals \cite{yassin_recent_2017}\cite{alarifi_ultra_2016}\cite{gezici_localization_2005}. In principle, the adopted tracking methods do not fundamentally differ from the ones employed in typical RF-based approaches, however, some special properties of UWB make it an attractive and promising solution. UWB communications are composed of very short pulses (shorter than 1ns) with a low duty cycle from 1 to 1000. The modulated signal is spread over multiple frequency bands and transmitted. Apart from communication applications, UWB is also considered for localization applications. The position estimation is typically performed through reference nodes of known positions through well-known techniques, such as Received Signal Strength (RSS), Angle of Arrival (AoA), Time of Arrival (ToA), and Time Difference of Arrival (TDoA). It has been shown that UWB can achieve high localization accuracy, mainly due to the decomposition of the multipath signal components in the channel's high bandwidth. However, the increased accuracy can be actually achieved only through time-based positioning methods, thanks to the signal's high time resolution, rather than RSS. The obvious drawback is that time-based techniques typically require good synchronization. In addition, UWB is usually of limited range, due to its high bandwidth, making it suitable mainly for short distance tracking.

It should be noted that all these target tracking approaches are based on distance measurements or estimation of the target position. Either relying on signal strength or angle or timing, estimation errors are inevitable, due to signal variations induced by shadowing, multipath fading or interference, or even due to the tracker's dead reckoning errors. In contrast, the proposed method does not involve distance measurements, taking advantage of the fact that estimating the exact target position is not required for the tracking process. The considered scenario does not aim at localizing the target, but staying in close range for the main reason of maintaining connectivity. This is achieved by exploiting the communication signal transmitted anyway by the target and considering its strength indicators in a differential manner. Such an approach does not require special communication equipment (such as directional antennas) nor transceivers' synchronization nor multiple reference points of known coordinates.

\subsection{Localization Strategies based on Swarm Controlling}
A promising approach for target localization that has recently risen is the use of robotic swarms. The control of drone swarms particularly has lately attracted significant interest, mainly due to the provided practicality and flexibility. In what follows in this subsection, we discuss a number of representative swarm controlling strategies for localization purposes. 

A network of drones that carry RF signal strength monitoring equipment is presented in \cite{liang_rf_2011}. The collected measurements are utilized to estimate distances based on a propagation model. The location of the radio transmitter is identified via trilateration of the resulted values. The authors in \cite{albert_combined_2019} have introduced a scheme which combines linear and non-linear programming to analyse the drones' movement constraints and perform both optimal and optimization control. Authors in \cite{liu_co-optimization_2018} proposed a technique for planning motion, according to which a number of drones are grouped together to perform target tracking in a collaborative manner that involves optimization of the monitoring process and the communication with a remotely located base station. The main goal is to ensure highly reliable connections to the base state, while at the same maximizing the collected sensor data. The provided evaluation results show that the process of optimizing transmissions is crucial for the integration of data and the precision of target positioning. In \cite{koohifar_receding_2017}, another promising approach is provided, according to which a swarm of drones perform collaborative localization of a radio transmitter using a control technique that is based on model prediction. The devices utilize the RSSI measurements of their radio receivers to identify the next optimal route adopting Receding Horizon Control (RHC). An Extended Kalman Filter is applied to generate the predicted parameters of the drones' motion, which are then tuned via using the D-optimality criterion. The authors in \cite{mavrommati_real-time_2018} adopt a similar strategy, which is also based on RHC to enable localization of multiple entities. The motion of the group of agents is determined based on the ergodic theory, where the information density distribution is adjusted by bearing-only measurements to track moving targets in real time.

\section{Overall IoT System Description}

\begin{figure}
 \centering
 \includegraphics[width=0.9\textwidth]{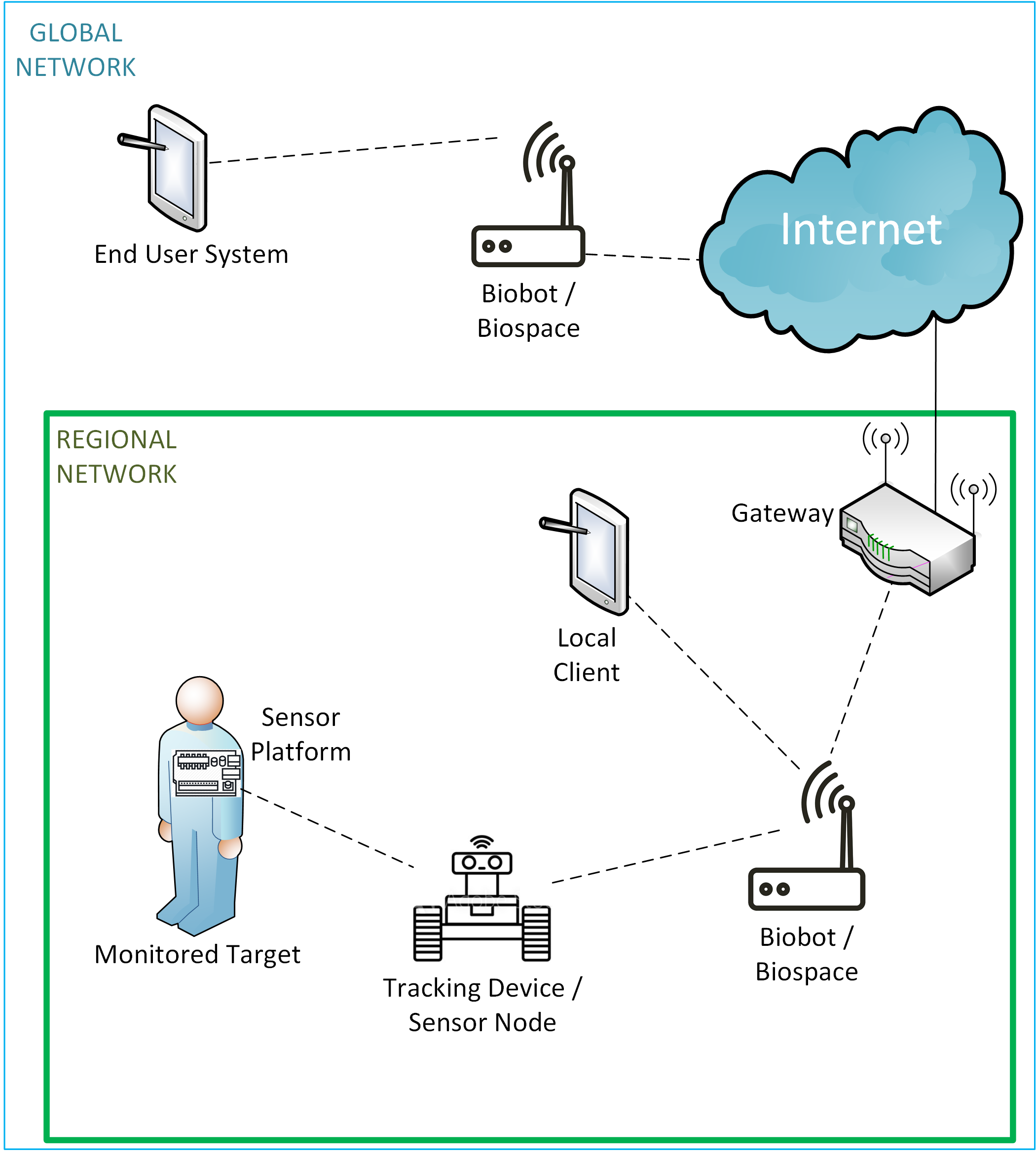}
 \caption{Architecture of the devised IoT System}
 \label{fig:architecture}
\end{figure}

The devised IoT system \cite{eleftherakis_architecting_2015} for monitoring mobile targets/individuals called eXtreme Sensor Network (XSN), comprises the "Global Network", and a number of "Regional Networks" that communicate through the former. An illustration of the main components of the system architecture is provided in Figure \ref{fig:architecture}.

\subsection{Global Network}
The agent-based distributed system is the part of the IoT architecture that realizes the global network, which allows efficient discovery of sensor data services and seamless access from remote locations over an unstructured distributed network. The main problem that is solved in that manner is the localization of the monitored target that carries the sensor platform and the reception of its data regardless of the exact regional network it is located in.
The implemented distributed system in the form of IoT middleware is based on the EDBO (Emergent Distributed Bio-Organization) architecture \cite{fortino_emergent_2013}
and was developed using JADEX \cite{braubach_developing_2012}. The components that fulfill the aforementioned requirements and realize distributed access and remote service discovery in the context of the considered architecture are the following:

\begin{itemize}
\item \textit{Biospace}: This is the platform that constitutes the basis for the creation of agents, which are able to serve sensor data.
\item \textit{Biobots}: These are the agents that access sensor nodes via RESTful requests, in order to provide information collected by sensor platforms. They are able to communicate in a distributed manner with each other for service discovery purposes.
\item \textit{End User Systems}: These are applications that access sensor services by communicating with the discovered Biobots. In the resulted IoT system, some Biobots are created in computing devices located in the same regional network with the sensor nodes, so that they have direct access to sensor data. From that point, information can be relayed over the global network. The end user systems are installed to user devices (such as tablets or smartphones) that remotely access sensor data or to remote servers that collect and process sensor information.
\end{itemize}

\subsection{Regional Network}
 Each local deployment which allows direct access to a followed mobile target's sensor information constitutes a Regional Network. Below, the system components which are included in such a deployment are described:
\begin{itemize}
\item \textit{Monitored Target}: This might be a person (for instance victim, patient or elderly transmitting vital signs) or any mobile target that has sensors attached (a Sensor Platform) to collect monitoring data.
\item \textit{Sensor Platform}: This is an embedded system with integrated sensors carried by the monitored target. It is also equipped with an energy efficient short-range wireless network interface which enables broadcasting sensor information. 
\item \textit{Tracking Device}: This an autonomous mobile robotic device which follows the monitored target with the purpose of maintaining proximity in order to gather and relay data generated by sensors or provide any type of assistance.
\item \textit{Sensor Node}: This is an embedded system with suitable wireless network interfaces to receive the signal broadcasted by the sensor platforms and then properly forward it, in the form of an entity which realizes mobile ad hoc network (MANET) routing. It ensures connectivity among multiple sensor nodes in the same regional network, but also connectivity with the Biobots of the Global Network. It may be fixed or carried by the tracking device and has sufficient processing capabilities to allow the creation of network services which enable client access to monitoring information. In the general sense and in the context of the regional network, it plays the role of the data sink.
%in the formed Wireless Sensor Network (WSN). 
\item \textit{Local Client}: This is an optional device handled by an end user located in the regional network to access sensor data. Since this entity lies in the region of the local network, it does not need to access the distributed service provided in the global network. It may directly query the Sensor Node to receive sensor data.
\item \textit{Gateway}: This is just a typical network device that plays the role of the gateway router for the regional network.
\end{itemize}

\section{The "Hot-Cold" Target Following Scheme}
% CONTENT COMMENTS:
% - Overall Concept
% - Algorithm Description
% -------------------------------------------------------
The main role of the robotic device in the system architecture is carrying a sensor node that is kept in range of the sensor platform. In that manner, the monitored individual can move freely. Of course, keeping the robot close can lead to many additional promising applications, such as delivering items to a person (e.g. medicine), providing assistive services (e.g. making emergency calls) or even keeping company (numerous studies have shown that robots could help elderly people as companion pets \cite{broekens_assistive_2009}). The primary goal is maintaining communication range; for that reason we have implemented an RF-based following scheme that solely uses the strength of the signal broadcasted by the sensor platform to estimate its location and move within range.

The RF-based following scheme uses the RSSI (Received Signal Strength Indicator) value. 
This is an indication of the signal power received by the sensor node and transmitted by the sensor platform. Our aim is the introduction of a simple and robust technique exploiting the RF signal which is anyway broadcasted by sensing devices for communication reasons. It should be highlighted that our main goal is maintaining communication range, not accurately locating the sensor platform. Hence, the developed RF-based following scheme aims at ensuring exactly that.

The concept behind this scheme is clear. As long as RSSI is not decreased, the robot keeps moving forward until a maximum RSSI threshold is reached, indicating that the robot is too close to the monitored person (we call this status "halt"). If RSSI decreases, then the robotic device rotates and moves towards a different direction, in order to avoid moving out of range. Due to the high unreliability of the wireless link, it is not safe to make final decisions each time there is a new RSSI reading, since it could just be a random deviation from the value that actually corresponds to distance. This is the first issue we cope with. The decisions related with RSSI change are based on statistic metrics of consecutive measurements, according to the following steps: 
\begin{enumerate}
\item 
The first step is the calculation of the mean RSSI value out of a number of samples stored in the Samples Window (SW). The window size is denoted by SWS (Samples Window Size).
\item
Next, after computing two mean values, we are considering the difference between the first and the second value. A positive difference (i.e. signal power increases) corresponds to the indication "Hot", whereas a negative difference (i.e. signal power decreases) corresponds to the indication "Cold".
\item
Lastly, a decision is made. If the indication is not "Cold", the robot moves forward, otherwise it rotates. The rotation intends to move the robot closer to the target.
\end{enumerate}
The "Hot-Cold" algorithm is presented in Algorithm 1.

\begin{algorithm}[H]
\caption{The Hot-Cold Algorithm}
 \begin{algorithmic}[1]
 \WHILE{$\text{Tracking-Following}$}
 \STATE $\text{SamplesAverage}\gets \emptyset$ 
 \STATE $j\gets 1$
 \WHILE{$j \leq 2$}
 \STATE $\text{RSSI}\gets \emptyset$
 \STATE $i\gets 1$
 \WHILE{$i \leq SWS$}
 \STATE $\text{IsHalt}\gets false$
 \STATE $\text{RSSI}[i]\gets \text{GetRSSI}$ 
 \STATE $\text{SamplesAverage}[j] \mathrel{+}= \text{RSSI}[i]$
 \IF{$\text{RSSI}[i] \leq HaltThreshold$}
 \IF{$i<SWS$}
 \STATE $\text{Move by robot step size}$
 \ENDIF
 \ELSE
 \STATE $\text{IsHalt}\gets true$
 \ENDIF
 \STATE $i++$
 \ENDWHILE
 \STATE $\text{SamplesAverage}[j] \mathrel{/}= SWS$
 \STATE $j++$
 \ENDWHILE
 \IF{$\text{IsHalt} = false$}
 \IF{$\text{SamplesAverage}[1] > \text{SamplesAverage}[2]$}
 \STATE $\text{Rotate}$
 \ENDIF
 \STATE $\text{Move by robot step size}$
 \ENDIF
 \ENDWHILE 
 \end{algorithmic}
\end{algorithm}

A crucial aspect of the introduced algorithm is the rotation angle. It is important to keep the algorithm simple and error tolerant. The whole scheme needs to exhibit advanced immunity to signal power variations, so that it is adequately robust to drive the robot close to the target. The exact localization is not significant; maintaining proximity is the highest priority. For these reasons, the main goal is to efficiently follow the target, while the rotation angle is fixed. We conclude on the optimal value of the rotation angle through a 3-stage analysis presented in the corresponding sub-sections below. The fourth subsection presents a convergence analysis which proves that by employing the introduced Hot-Cold algorithm, the robot reaches the followed target in finite number of steps.

\subsection{Geometrical Analysis of Rotation Angle}
The first stage of this analysis focuses on the geometric properties of the proposed target following scheme. The objective here is to estimate a range of rotation angles which rapidly move the robotic device closer to the area where the target is most probably located.

\begin{figure}[!t]
\centering
\subfloat[Rotation angle: 120 degrees]{\includegraphics[width=.6\textwidth]{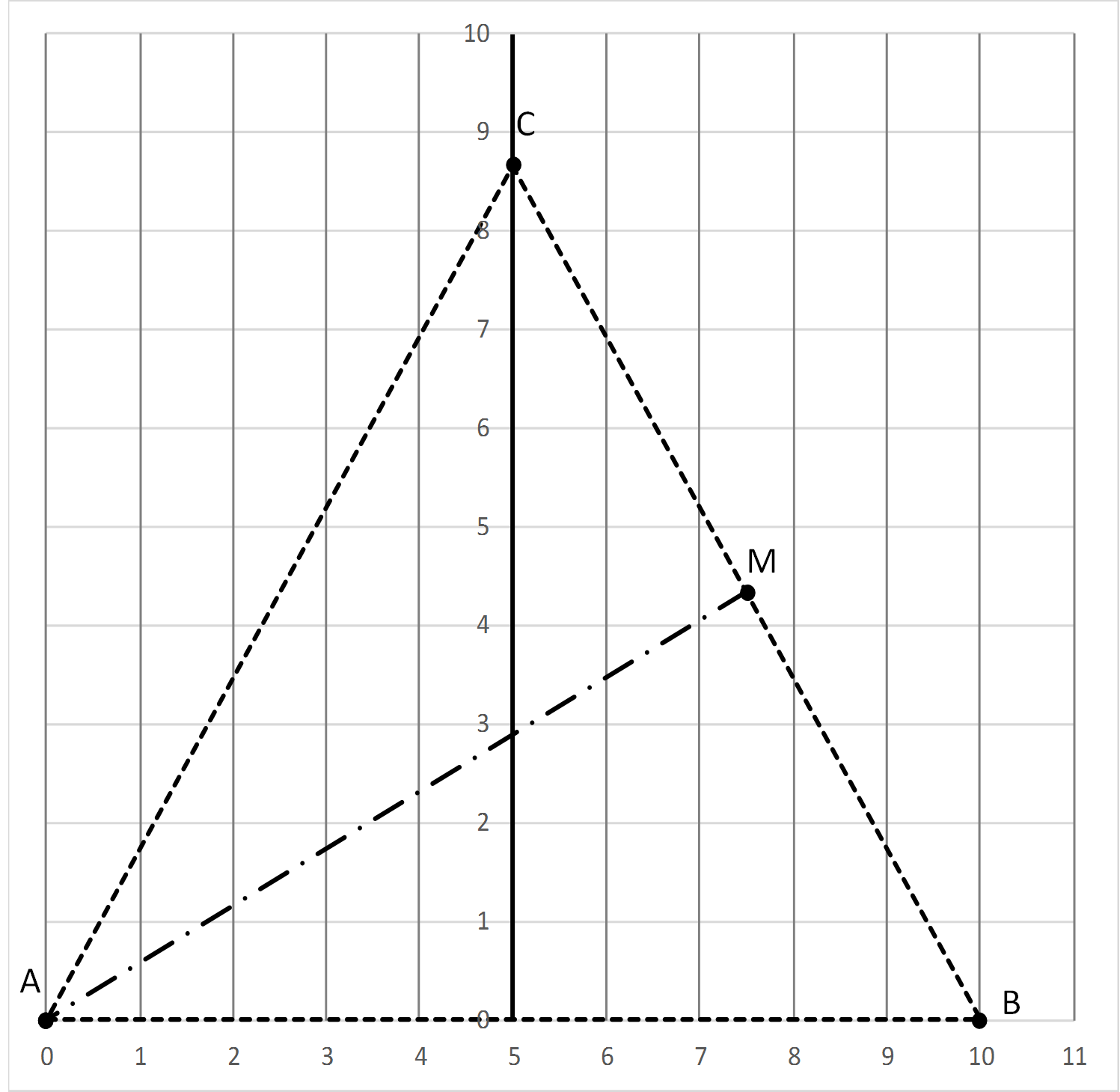}%
\label{fig:geometric_analysis_120}}
\hfil
\subfloat[Rotation angle: 144 degrees]{\includegraphics[width=.6\textwidth]{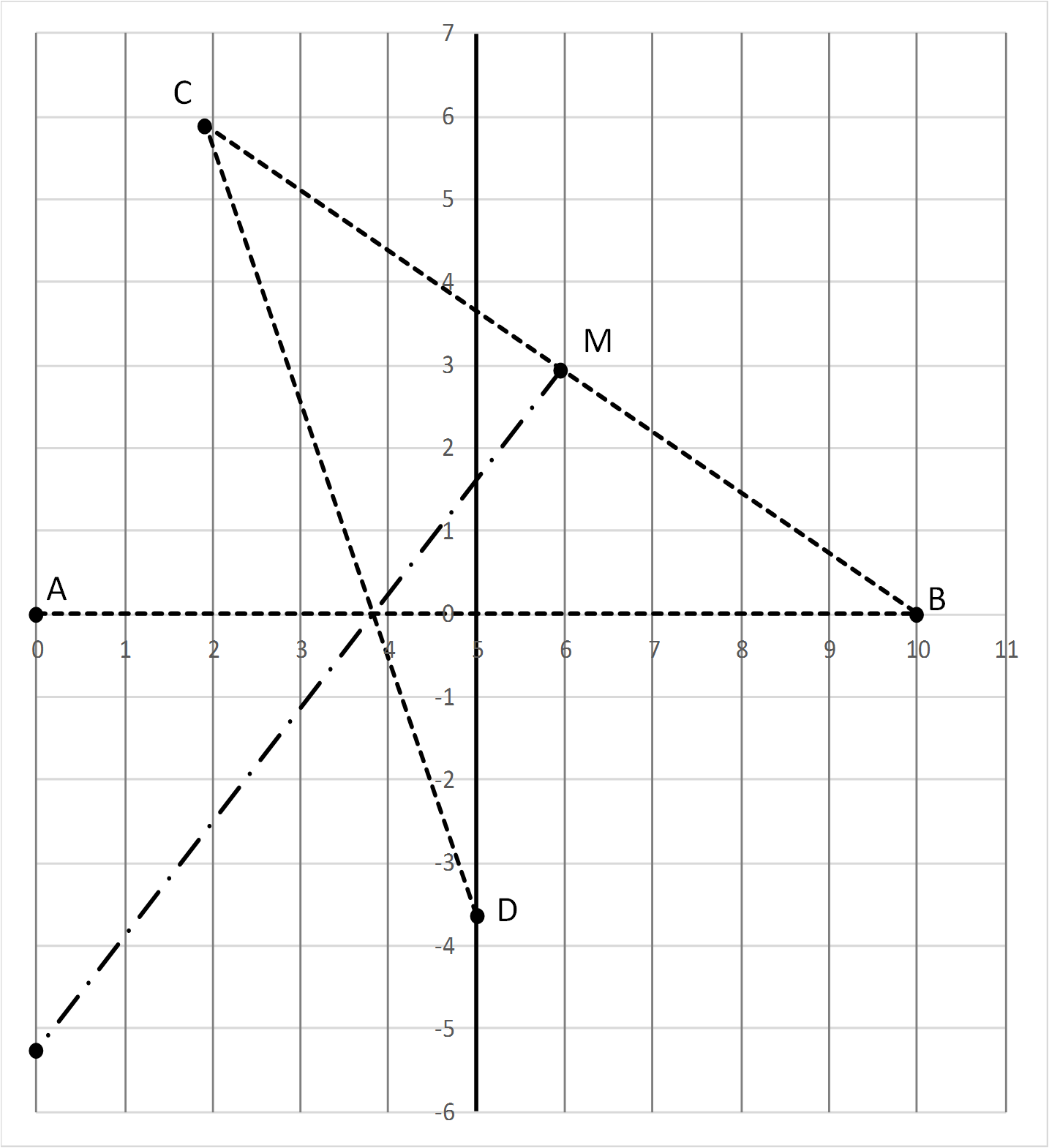}%
\label{fig:geometric_analysis_144}}
\caption{Three consecutive steps with rotations at angle of a) 120 degrees and b) 144 degrees}
\label{fig:geometric_analysis}
\end{figure}

We consider the scenario where the robot moves from point A to point B by one step equal to 10 distance units, as presented in Figure \ref{fig:geometric_analysis_120}. The starting robot position is A and a step later it is located at position B. Assuming that at that point the Hot-Cold algorithm deduces that the robot has moved away from the target, a rotation should be performed. In this example, the rotations are considered to be counter-clockwise, without loss of generality. The fact that the target is closer to A than B indicates that it is located in the area left from the vertical line bisector of the segment AB (note that the depicted y-axis lies on this bisector). In order to reach this area, the rotation angle needs to be higher than 120 degrees and lower than 270 degrees. The figure illustrates rotation by 120 degrees, which positions the robot after its second step at point C. Assuming again that at point C the robot is further from the target than it was at point B, a new rotation will take place. The target should be now positioned on the lower half of the area divided by the vertical line bisector AM, where M is the midpoint of the line sector AB. It is noted that point A lies on this bisector, when the robot rotates by 120 degrees counter-clockwise, as illustrated in Figure  ~\ref{fig:geometric_analysis_120}. Taking also into account the observation of the first step which dictated that the target is on the left side of the y-axis, it is proven that the target is located in an area that is accessible by performing fixed angle rotation between 120 to 144 degrees. Specifically, this area, where the target is positioned, lies under the line defined by the bisector AM and left from the line defined by the y-axis. When rotating by 120 degrees, the robot will be positioned after the third step back to point A, which is at the borderline of the target area.

%\begin{figure}
% \includegraphics[width=\textwidth]{rotation-analysis_120.png}
% \caption{Three consecutive steps with rotations at 120 degrees}
% \label{fig:geometric_analysis_120}
%\end{figure}

The graphical depiction of the the 3-step movement when rotating by 144 degrees is provided in Figure  \ref{fig:geometric_analysis_144}. It can be seen that after three steps, the robot reaches position D, which is on the right border of the target area that lies under the line bisector intersecting point M and left from the line defined by y-axis. Please note that in the case of 120-degrees rotation illustrated in Figure \ref{fig:geometric_analysis_120}, point D overlaps with point A. 
%\begin{figure}
% \includegraphics[width=\textwidth]{rotation-analysis_144.png}
% \caption{Three consecutive steps with rotations at 144 degrees}
% \label{fig:geometric_analysis_144}
%\end{figure}
In conclusion, the geometrical analysis reveals that the fixed rotation angle has to be higher than 120 and lower than 144 degrees for the robot to reach the target area in the minimum number of steps, when adopting Hot-Cold target following. Obviously, rotation angles out of this range could eventually drive the robot in the target area, however, on average more steps would be required, whereas the objective is to reach the target as fast as possible.

\subsection{Numerical Analysis of Rotation Angle}
The second stage of the analysis in the effort to identify the optimal rotation angle for efficient following via the Hot-Cold algorithm focuses on calculating the number of fixed rotations required to reach a target point at any angle. Specifically, the objective is to identify the rotation angle $\phi$, which minimizes the average number of rotations required ($\overline{\omega}$) to reach any target angle $\theta_{i}$ within a deviation $\pm\varepsilon$.

The mathematical expression that relates the aforementioned variables is shown in Eq.(\ref{eq:geometric_analysis_A}).
\begin{equation}
 \theta_{i} - \varepsilon \leq (\omega_{i}\times\phi)mod360 \leq \theta_{i} + \varepsilon
 \label{eq:geometric_analysis_A}
\end{equation}
Solving Eq.(\ref{eq:geometric_analysis_A}) for $\omega$ leads to Eq.(\ref{eq:geometric_analysis_B}).
\begin{IEEEeqnarray}{C}
 \theta_{i} - \varepsilon \leq \omega_{i}\phi-360\kappa_{i} \leq \theta_{i} + \varepsilon \Rightarrow \nonumber\\
 \theta_{i} - \varepsilon + 360\kappa_{i} \leq \omega_{i}\phi \leq \theta_{i} + \varepsilon + 360\kappa_{i} \Rightarrow \nonumber\\
  \frac{\theta_{i} - \varepsilon + 360\kappa_{i}}{\phi} \leq \omega_{i} \leq \frac{\theta_{i} + \varepsilon + 360\kappa_{i}}{\phi}
 \label{eq:geometric_analysis_B}
\end{IEEEeqnarray}
where $\kappa_{i}$ is the lowest non-negative integer that makes $\omega_{i}$ integer. The optimization problem is formulated as follows:
\begin{equation}
 \phi = \argmin{\overline{\omega}} \textnormal{ , } \forall{i}\in[1,{n}]\ \textnormal{ , where } {n}\in \mathbb{Z^+}
\label{eq:argmin}
\end{equation}
s.t.
\begin{equation}
 \phi,\theta_{i},\varepsilon \in [0,360)
\label{eq:condition_i}
\end{equation}
\begin{equation}
 \phi \in \mathbb{Z}
\label{eq:condition_ii}
\end{equation}
\begin{equation}
 \omega_{i} \in \mathbb{Z^*}
\label{eq:condition_iii}
\end{equation}
\begin{equation}
 \kappa_{i} \in [0,\phi] \land \kappa_{i} \in \mathbb{Z}
\label{eq:condition_iv}
\end{equation}

In order to identify $\omega_{i}$, we solve Eq.(\ref{eq:geometric_analysis_B}) in a numerical approach for all integer values of $\theta_{i}$ and $\phi$, and different values of $\varepsilon$. Given that $\kappa_{i}\in[0,\phi]$, consecutive integer values of $\kappa_{i}$ are tested in each iteration until the first solution of Eq.(\ref{eq:geometric_analysis_B}) is found. The optimal $\phi$ is the one which yields the lowest: 
\begin{equation}
 \overline{\omega}= \frac{\sum_{i=1}^{n}{\omega_{i}}}{n}
\label{eq:condition_v}
\end{equation}
The procedure that provides numerical solution to the described problem through iterative trials was developed and executed in MATLAB. In more detail, we tested all rotation angles from 121 up to 143 degrees according to the findings of the geometrical analysis (120 and 144 degrees angles are borderline cases, hence, non-optimal). It is noted that only angles of integer degrees are considered; further subdividing angles makes no difference, since the potential benefits would be minimal and in a real scenario a robotic following device could not make that accurate turns anyway. For each rotation angle ($\phi$) we compute the number of rotations required ($\omega_{i}$) to reach any target angle ($\theta_{i}$) from 0 to 359 degrees within a deviation ($\pm\varepsilon$) from 0 to 30 degrees. 

\begin{figure}
 \centering
 \includegraphics[width=0.9\textwidth]{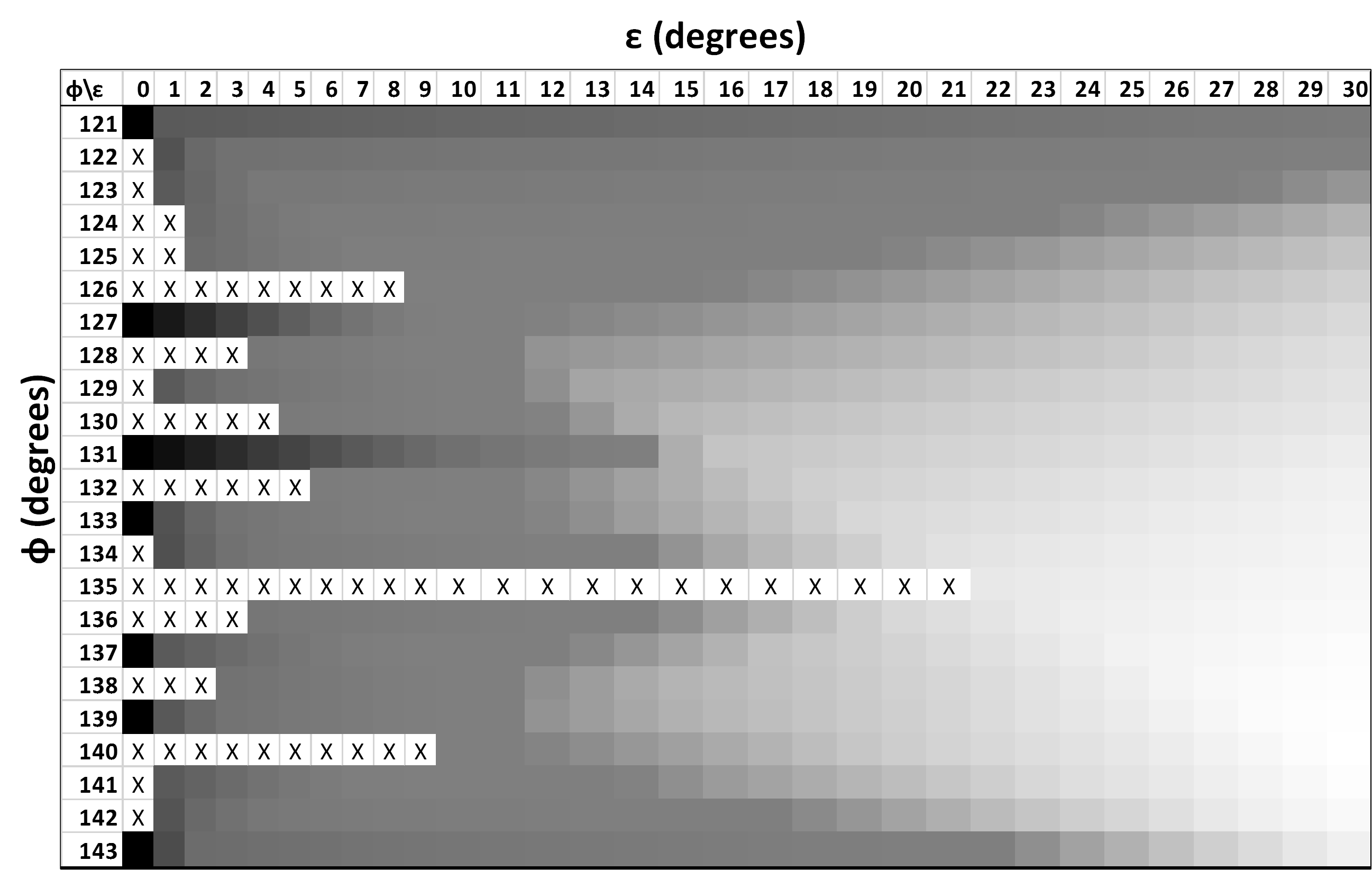}
 \caption{Heatmap of average number of rotations required for different rotation angles ($\phi$) and target angle deviations ($\pm\varepsilon$) in degrees | Legend: Lightest is 3, Darkest is 179.5, 'X' is invalid}
 \label{fig:rotations_required_simple}
\end{figure}

\begin{figure}
 \centering
 \includegraphics[width=0.9\textwidth]{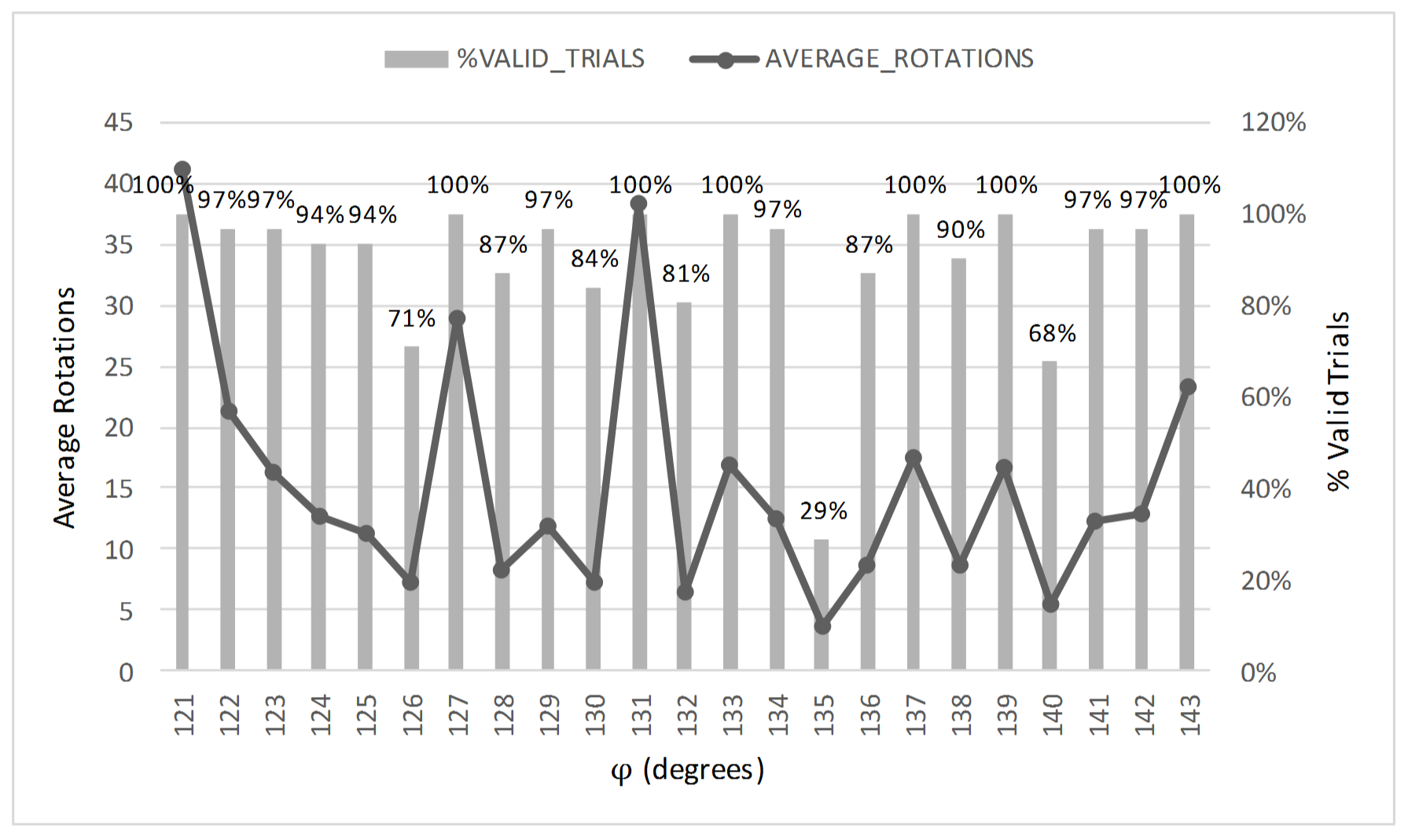}
 \caption{Overall average number of rotations required and percentage of valid trials against different rotation angles ($\phi$)}
 \label{fig:rotations-valid_trials}
\end{figure}

Figure  \ref{fig:rotations_required_simple} provides a heatmap of $\overline{\omega}$ values that have been derived by averaging over all 360 values of $\theta_{i}$. As expected, high $\varepsilon$ ensures low average number of rotations required (lighter regions). Cells with "X" represent cases where it was impossible to reach some target angles ($\theta_{i}$) within the respective deviation ($\varepsilon$). The optimal rotation angle should require low number of rotations to reach a large number of target angles within small deviation. In order to conclude on the value of this angle according to the specific criteria, we have further averaged the $\overline{\omega}$ values over each rotation angle ($\phi$) and plotted them along with the percentage of valid trials in Figure  \ref{fig:rotations-valid_trials}. It can be deduced that the lowest mean number of rotations (16.78) with no invalid trials is achieved when $\phi$ equals 139 degrees.    

\subsection{Exhaustive-Simulation Analysis of Rotation Angle}
The third stage of the analysis for identifying the optimal rotation angle that would efficiently drive the robotic device close to the target involves exhaustive simulations in MATLAB. Specifically, the objective of this final part of the analysis is to compute the number of required steps taken by the robotic device to approach a fixed target when adopting the Hot-Cold algorithm principles. In each configuration, the target is placed $\rho$ distance units away from the robot's starting position and at a direction of $\beta$ degrees. The followed approach is actually exhaustive; simulations are executed for all integer values of $\beta$ ranging from 0 to 359 degrees and $\rho$ ranging from 10 to 100 distance units. Each simulation is terminated when the robotic device approaches the target within $\tau$ distance units; all integer values from 1 to 10 are tested. It is noted that a distance unit is set equal to the length of one step. Based on the Hot-Cold concept, the robot rotates at a fixed angle every time its new position is at a higher distance from the target than its previous position. Simulations are performed for all integer rotation angles ($\phi$) from 121 to 143 degrees, with the total number of executed simulations just for this last stage of the analysis reaching 7,534,800. 

\begin{figure}
 \centering
 \includegraphics[width=\textwidth]{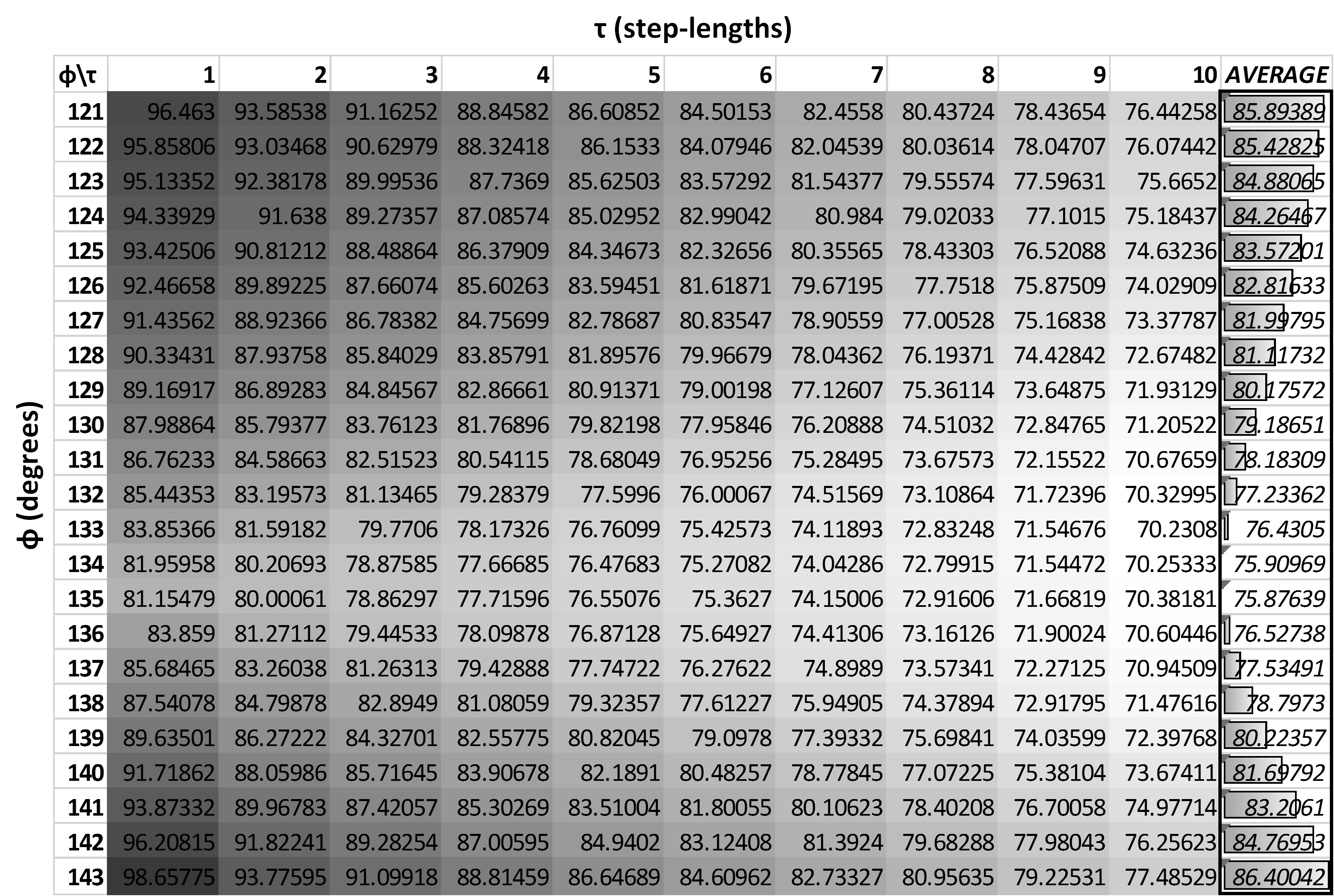}
 \caption{Mean number of steps required against rotation angle ($\phi$) in degrees and halt distance from target ($\tau$) in step-lengths}
 \label{fig:steps_required}
\end{figure}

The findings are depicted in the heatmap of Figure  \ref{fig:steps_required}. For each combination of $\phi$ and $\tau$, we calculate the mean number of steps considering all $\rho$ and $\beta$ values; the results are shown in the cells of the heatmap. As expected, the more relaxed the termination condition is (high $\tau$ values), the fewer steps are required (lighter regions). Regarding the optimal rotation angle, a clear pattern is revealed, especially when looking at the overall averages presented in the last column of the figure. It is evident that the closer a rotation angle is to 135 degrees, the fewer steps are required to reach the target. The minimum number of 75.87639 averaged steps is achieved for $\phi$ exactly equal to 135 degrees. 

Conclusively, the conducted 3-stage analysis shows that the optimal rotation angle for a robotic device adopting the introduced Hot-Cold algorithm to approach a target as fast as possible is in the range of 135 to 139 degrees. Considering the inevitable declination from the set rotation angle of a robotic vehicle, configuring it to the median value of 137 degrees is a safe choice.    

\subsection{Convergence Analysis}
Following the 3-stage analysis for the determination of the optimal rotation angle, in this subsection a convergence analysis is presented, which was conducted to prove whether the Hot-Cold algorithm theoretically ensures target approaching within a finite number of steps. We break down this analysis in two parts, Hot mode and Cold mode, demonstrating that in both modes the introduced algorithm manages to converge robot's position close to target's position. 
\subsubsection*{Lemma 1}
A robotic device which performs target following using the Hot-Cold algorithm always approaches the target while in Hot mode, given that their horizontal distance is greater than half of the robot's step, considering no signal fading, SWS equal to 1, and random walk as the target's mobility model.
\subsubsection*{Proof 1}
Regarding the target's movement, since it employs random walk, in every step it changes its distance ($d$) from the robot in a uniform manner, with a mean value of 0.

Focusing on the robot's movement, being in Hot mode means that the current distance $d$ is lower or equal to the corresponding distance during the previous step. It is noted that the two distances are directly comparable using the respective RSSI values, since for this theoretical analysis we assume that signal strength is only affected by propagation attenuation, not fading of any kind.

\begin{figure}
 \centering
 \includegraphics[width=.4\textwidth]{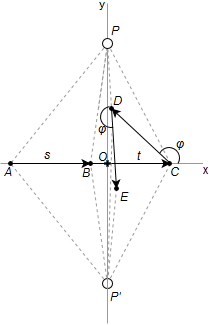}
 \caption{General tracking scenario, representing robot's 4 consecutive steps, transiting from Hot mode ($AB, BC$) to Cold mode ($CD, DE$)}
 \label{fig:convergence_fig}
\end{figure}

Figure  \ref{fig:convergence_fig} illustrates a general tracking scenario, where the robot performs four steps ($AB=BC=CD=DE=s$) and is located in five consecutive points (starting point $A$, ending point $E$). Covering the general model, we consider two alternative locations (at opposite sides symmetrical to the horizontal axis) for the target: $P$ and $P'$. The application of the Pythagorean theorem yields the following equations for robot’s first step:

\begin{equation}
 AP^2=PO^2+AO^2
 \label{eq:convergence_analysis_i}
\end{equation}
\begin{equation}
 BP^2=PO^2+BO^2=PO^2+(AO-AB)^2
 \label{eq:convergence_analysis_ii}
\end{equation}
\begin{IEEEeqnarray}{C}
 BP \leq AP \Rightarrow BP^2 \leq AP^2 \Rightarrow \nonumber\\ PO^2+(AO-AB)^2 \leq PO^2+AO^2 \Rightarrow \nonumber\\
 AO \geq s/2
 \label{eq:convergence_analysis_iii}
\end{IEEEeqnarray}
Eq.(\ref{eq:convergence_analysis_iii}) proves that the robotic device stays in Hot mode and moves forward approaching the target (current distance not greater than the previous one) as long as their horizontal distance ($AO$) just before the current step is not shorter than the robot's half step size ($s/2$). At the point this condition ceases to hold (point $B$), the robot increases distance $d$ with its immediate next step (point $C$), so transits to Cold mode. It is noted that due to symmetry the same also holds when the target is positioned at $P'$.

\subsubsection*{Lemma 2}
A robotic device which performs target following using the Hot-Cold algorithm approaches in Cold mode the target right after the first rotation, when their signed vertical distance is higher than $0.7304s-1.0649t$ ($t$ is their horizontal absolute distance, $s$ is the robot step size and $t>\tfrac{s}{2}$) or otherwise right after the second rotation, considering no signal fading, SWS equal to 1, and random walk as the target's mobility model.

\subsubsection*{Proof 2}
Using as reference Figure  \ref{fig:convergence_fig}, we now focus on the case that the robot transits to Cold mode when it moves to point $C$. The condition for this transition is $CP>BP$ (or equivalently $CP'>BP'$), which yields $CO>BO$. Given that $BC=s$, the latter condition holds when $t=OC>s/2$. According to the principles of the Hot-Cold algorithm, we consider that the robot rotates at point $C$ by $\phi=137^{\circ}$ counterclockwise, without loss of generality (in case of clockwise rotation, points $P$ and $P'$ can be just considered exchanged). Right after the rotation and robot's movement by one step-size ($s$), its new location is $D$ and there can be two cases regarding its distance from the target: i) it has been decreased (e.g. $DP<CP$) or ii) it has not been decreased (e.g. $DP'>CP'$). Hence, this analysis initially focuses on identifying the relation between the target's y-coordinate (denoted by $r_1$) and the robot's ability to approach right after its first rotation (point $D$). Specifically, we estimate the $r_1$ threshold which ensures that the considered distance after rotating becomes smaller. This part of the problem is formulated and solved as follows, where $t>\tfrac{s}{2}$:
\begin{IEEEeqnarray}{C}
 CP > DP \Rightarrow CP^2 > DP^2 \Rightarrow \nonumber\\
 r_1^2+t^2>(t+s\cos(\phi))^2+(s\sin(\phi)-r_1)^2 \Rightarrow \nonumber\\
 0>s^2(\cos^2(\phi)+\sin^2(\phi))+2ts\cos(\phi)-2r_1s\sin(\phi) \Rightarrow \nonumber\\
 2r_1\sin(\phi)>s+2t\cos(\phi) \xRightarrow[]{\text{$0 \leq \phi \leq \pi$}} \nonumber\\
 r_1>\tfrac{1}{2}s\csc(\phi)+t\cot(\phi) \xRightarrow[]{\text{$\phi=137^{\circ}$}} \nonumber\\
 r_1>0.7304s-1.0649t
 \label{eq:convergence_analysis_iv}
\end{IEEEeqnarray}

Next, this analysis focuses on the condition for approaching the target right after the second rotation, which requires that at the first rotation the robot increased its distance, hence, it remained in Cold mode. In Figure  \ref{fig:convergence_fig}, this is the case when the target is located at point $P'$. Following an approach similar to the above, it holds (where $t>\tfrac{s}{2}$):
{\allowdisplaybreaks
\begin{IEEEeqnarray}{C}
 DP > EP \Rightarrow DP^2 > EP^2 \Rightarrow \nonumber\\
 (t+s\cos(\phi))^2+(s\sin(\phi)-r_2)^2> \nonumber\\
 (t+s\cos(\phi)+s\cos(2\phi))^2+(s\sin(\phi)+s\sin(2\phi)-r_2)^2 \Rightarrow \nonumber\\
 s+2t\cos(\phi)-2r_2\sin(\phi)> \nonumber\\
 s(\cos(\phi)+\cos(2\phi))^2+2t(\cos(\phi)+\cos(2\phi))+ \nonumber\\
 s(\sin(\phi)+\sin(2\phi))^2+2r_2(\sin(\phi)+\sin(2\phi)) \Rightarrow \nonumber\\
 2r_2\sin(2\phi)>4s\cos^2(\tfrac{\phi}{2})+2t\cos(2\phi)-s \xRightarrow[\tfrac{3\pi}{2} \leq \phi \leq 2\pi]{\text{$\tfrac{\pi}{2} \leq \phi \leq \pi$}} \nonumber\\
 r_2<\dfrac{4s\cos^2(\tfrac{\phi}{2})+2t\cos(2\phi)-s}{2\sin(2\phi)} \xRightarrow[]{\text{$\phi=137^{\circ}$}} \nonumber\\
 r_2<0.2294s-0.0629t
 \label{eq:convergence_analysis_v}
\end{IEEEeqnarray}}

From Eq.(\ref{eq:convergence_analysis_iv}) and Eq.(\ref{eq:convergence_analysis_v}) it derives that the $r_1$ threshold is always lower than the $r_2$ threshold, given that $t>\tfrac{s}{2}$, which is always true. This means that if the robot moves away form the target after the first rotation, it will definitely approach it after the second rotation, transiting from Cold to Hot mode. Furthermore, following the same analytical method, it is shown that in case of two required rotations, the robot-target distance (illustrated by $EP'$ in Figure  \ref{fig:convergence_fig}) is eventually smaller than the original distance before any rotations ($CP'$). Specifically, for the general case of considering target's position as $P$, the problem is formulated as follows, where $r_3$ is the target's y-coordinate and $t>\tfrac{s}{2}$:
{\allowdisplaybreaks
\begin{IEEEeqnarray}{C}
 CP > EP \Rightarrow CP^2 > EP^2 \Rightarrow \nonumber\\
 r_3^2+t^2> \nonumber\\
 (t+s\cos(\phi)+s\cos(2\phi))^2+(s\sin(\phi)+s\sin(2\phi)-r_3)^2 \Rightarrow \nonumber\\
 0>s(\cos(\phi)+\cos(2\phi))^2+2t(\cos(\phi)+\cos(2\phi))+ \nonumber\\
 s(\sin(\phi)+\sin(2\phi))^2-2r_3(\sin(\phi)+\sin(2\phi)) \Rightarrow \nonumber\\
 r_3(\sin(\phi)+\sin(2\phi)> \nonumber\\
 2s\cos^2(-\tfrac{\phi}{2})+t(\cos(\phi)+\cos(2\phi)) \xRightarrow[]{\text{$\phi=137^{\circ}$}} \nonumber\\
 r_3<2.1251t-0.8646s
 \label{eq:convergence_analysis_vi}
\end{IEEEeqnarray}}

From Eq.(\ref{eq:convergence_analysis_iv}) and Eq.(\ref{eq:convergence_analysis_vi}) it derives that the $r_1$ threshold is always lower than the $r_3$ threshold, given that $t>\tfrac{s}{2}$, which is always true. Thus, it is proven that starting from the position where the robot enters the Cold mode (point $C$ in Figure  \ref{fig:convergence_fig}), it will always approach the target either right after the first $137^{\circ}$ rotation in case their signed vertical distance is higher than $0.7304s-1.0649t$ or right after its second rotation, otherwise.
 
\section{System Evaluation}
% CONTENT COMMENTS:
% - Simulation-based evaluation of the patient tracking scheme
% -- Comparison with Trilateration
% -- Results about parameters impact/optimization
% - Prototype-based evaluation [Including UML diagram of robot]
% ---------------------------------------------------------------
The evaluation of the devised system is based on a dual approach. Initially, we focus on the introduced Hot-Cold target following algorithm, which is thoroughly evaluated in a simulation-based manner. Then, a prototype is implemented, which is tested in controlled laboratory conditions. 

\subsection{Simulation-based Evaluation of Target Following Scheme}
%(based on Tracking-Figures created on 27/9/16)
In order to thoroughly evaluate the main focus of this work, which is the introduced Hot-Cold RF-based target following scheme, a simulator \cite{tlagkas_tlagkashot-cold_simulation_2020} was developed in the Processing Integrated Development Environment \cite{fry_visualizing_2007}. There are two main objectives of the conducted simulations: i) identify the optimal SWS (Samples Window Size) values and ii) evaluate Hot-Cold performance by comparing it against a reference target following scheme. The set values of the main simulation parameters are shown in Table \ref{tab:sim_param}. The direction change in the simulator takes place in two simulation cycles, that is 1 sec duration. At this point, it is clarified that the purpose of the simulations is to conduct comparison-based evaluation, which is successfully achieved by setting the exact same parameter values to the different simulation settings which are compared against each other. It is also noted that radio propagation modeling is based on the log-distance path loss model with log-normal shadowing, since it is widely accepted and generic enough to simulate various environments \cite{erceg_empirically_1999}. Path loss at the reference distance is estimated according to Friis formula \cite{rappaport_wireless_2001}, resulting in the following equation for estimating signal path loss:
\begin{equation} 
    \label{eq:pathloss}
    \begin{split}
        PL = 10n\log{d} + 20\log{f} + 20\log{\tfrac{4\pi}{c}} + X_{g}
    \end{split}
\end{equation}
\noindent where $d$ is the transmitter-receiver distance, $f$ is the central frequency, $c$ is the speed of light, $X_{g}$ is a Gaussian random variable with mean $\mu = 0$ and standard deviation $\sigma$ modelling slow fading due to mobility/shadowing. It should be noted that the correlation between the received signal strength and the distance is verified in multiple studies including the experiments and regression analyses performed in the context of WINNER I \cite{meinila20052003} and WINNER II \cite{kyosti1winner} projects on wireless channel modeling. However, it is undeniable that this correlation is degraded by the presence of any form of noise (such as fading and interference). Driven by this fact, the proposed technique avoids the direct computation of distances based on RSSI, rather it utilizes indications of RSSI changes to roughly deduce whether the robotic device approaches the target or not. 

\begin{table}[!t] 
\renewcommand{\arraystretch}{1.3} 
\caption{Simulation Parameters} 
\label{tab:sim_param} 
\centering 
\begin{tabular}{c|c} 
\hline 
\bfseries Simulation Parameter & \bfseries Value\\ 
\hline\hline
Simulation duration & 1000 sec\\ 
\hline 
Executions per simulation & 5\\ 
\hline 
Simulation space & 100$\times$100 $m^2$\\ 
\hline 
Robot speed & 7.2 km/h (default)\\ 
\hline 
Target speed & 3.6 km/h\\ 
\hline 
Target mobility pattern & Random waypoints\\ 
\hline 
Halt distance & 3 m\\ 
\hline 
Broadcast interval / Simulation cycle period & 0.5 sec\\ 
\hline 
Target TX power & 0 dBm\\ 
\hline 
Robot RX sensitivity & -94 dBm\\ % changed from -92 
\hline
Target TX antenna gain & 0 dBi\\ % changed from 2
\hline
Robot RX antenna gain & 2 dBi\\ 
\hline
Path loss exponent & 2.8\\ 
\hline
Signal frequency & 2.4 GHz\\ 
\hline
\end{tabular} 
\end{table}

\begin{figure}
 \centering
 \includegraphics[width=\textwidth]{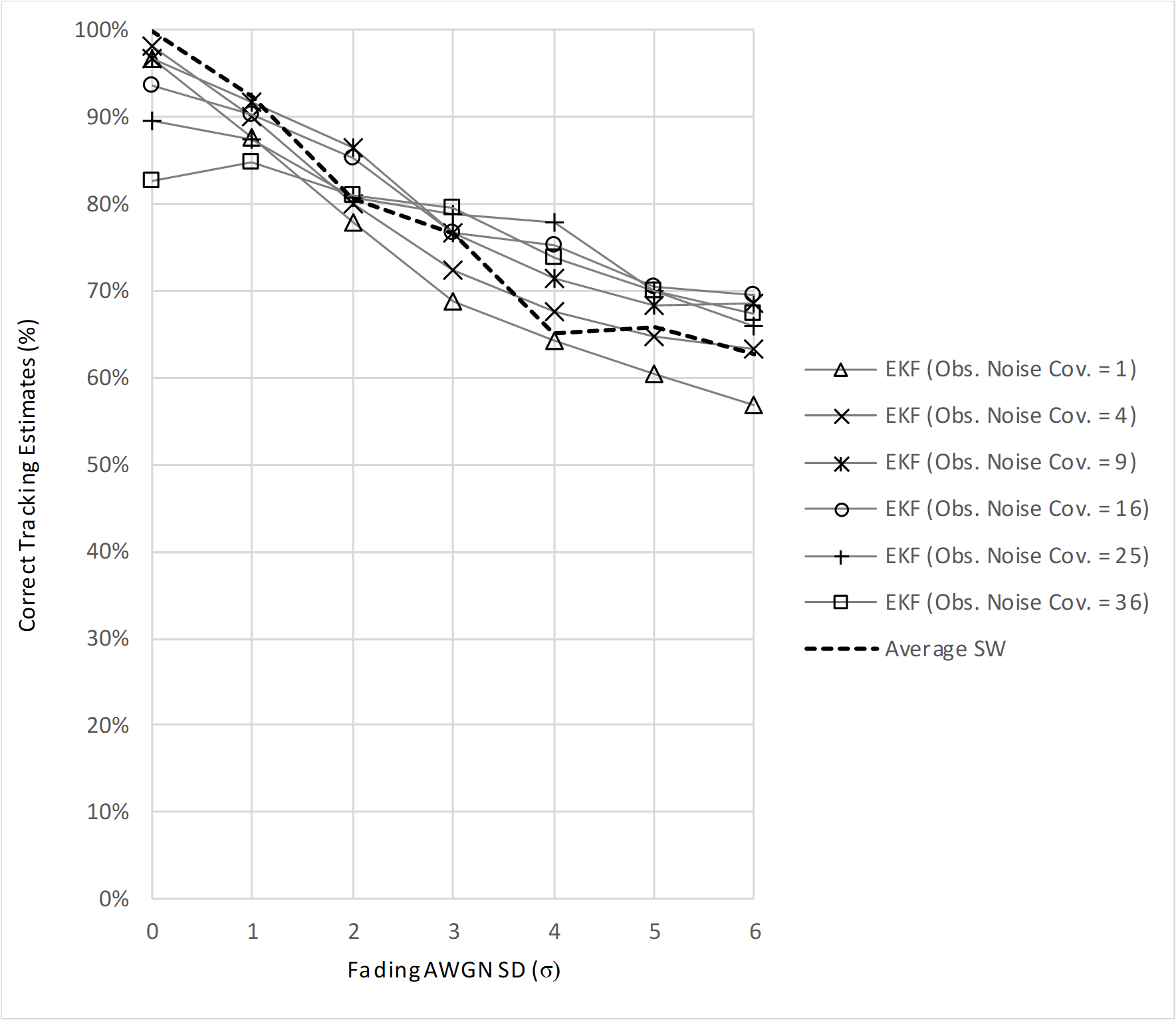}
 \caption{Correct tracking estimates of the adopted SW averaging technique compared with EKF for different values of observation noise covariance.}
 \label{fig:EKF_vs_AVG}
\end{figure}

Before proceeding with the performance evaluation, we first compare the efficiency of the adopted technique of averaging RSSI values within the Samples Window (SW) against the dominant Extended Kalman Filtering (EKF) technique. We have chosen EKF as a reference noise cancellation algorithm, on the grounds that it is probably the most widely accepted scheme for nonlinear state estimation in navigation systems, such as the Global Positioning System \cite{4378854}. In our case, EKF was applied on the observed RSSI values that vary as the target moves and the robot follows it, which obviously constitutes a nonlinear model. The only observable metric is the received RSSI, while the estimated state is the true RSSI (i.e. relieved from noise), with both values being scalar. The EKF observation model as well as the process noise covariance were set to 1. On the other hand, the observation noise covariance, which optimally reflects the standard deviation of the Gaussian noise experienced by the receiver, is set to varying values given that the exact shadowing effect is unknown. Figure  \ref{fig:EKF_vs_AVG} presents the percent of correct tracking estimates (i.e. Hot or Cold) made when we replace in the introduced algorithm the SW averaging with EKF for different values of observation noise covariance. The dashed line depicts the respective results when we average the values within the SW, as described in Algorithm 1. It becomes evident that in the general case, the two techniques perform similarly in the context of the Hot-Cold algorithm. Of course, it should be clarified that EKF and its variations can be very promising, since they are highly adjustable and can be tuned for specific mobility and signal propagation models. It is undeniable that EKF exhibits its great potential when fusing the readings from multiple sensors. However, since in the considered environment only RSSI readings are assumed to be available and there can be no safe inference about the target's mobility model, the SW averaging technique is adopted. Nevertheless, it is clear that Hot-Cold is modular enough to allow replacements of its RSSI estimating component with any suitable filtering technique, such as EKF.

The key performance indicators that are used for evaluation purposes are the following: i) Average Distance: The distance between the robot and the target averaged over the simulation duration. Lower values indicate better performance. ii) Cycles in Range: Simulation cycles during which the robot stays in the communication range of the target. Higher values indicate better performance. iii) Cycles in Halt: Simulation cycles during which the robot freezes, due to short (halt) distance from the target. Higher values indicate better performance.

\begin{figure}
 \centering
 \includegraphics[width=\textwidth]{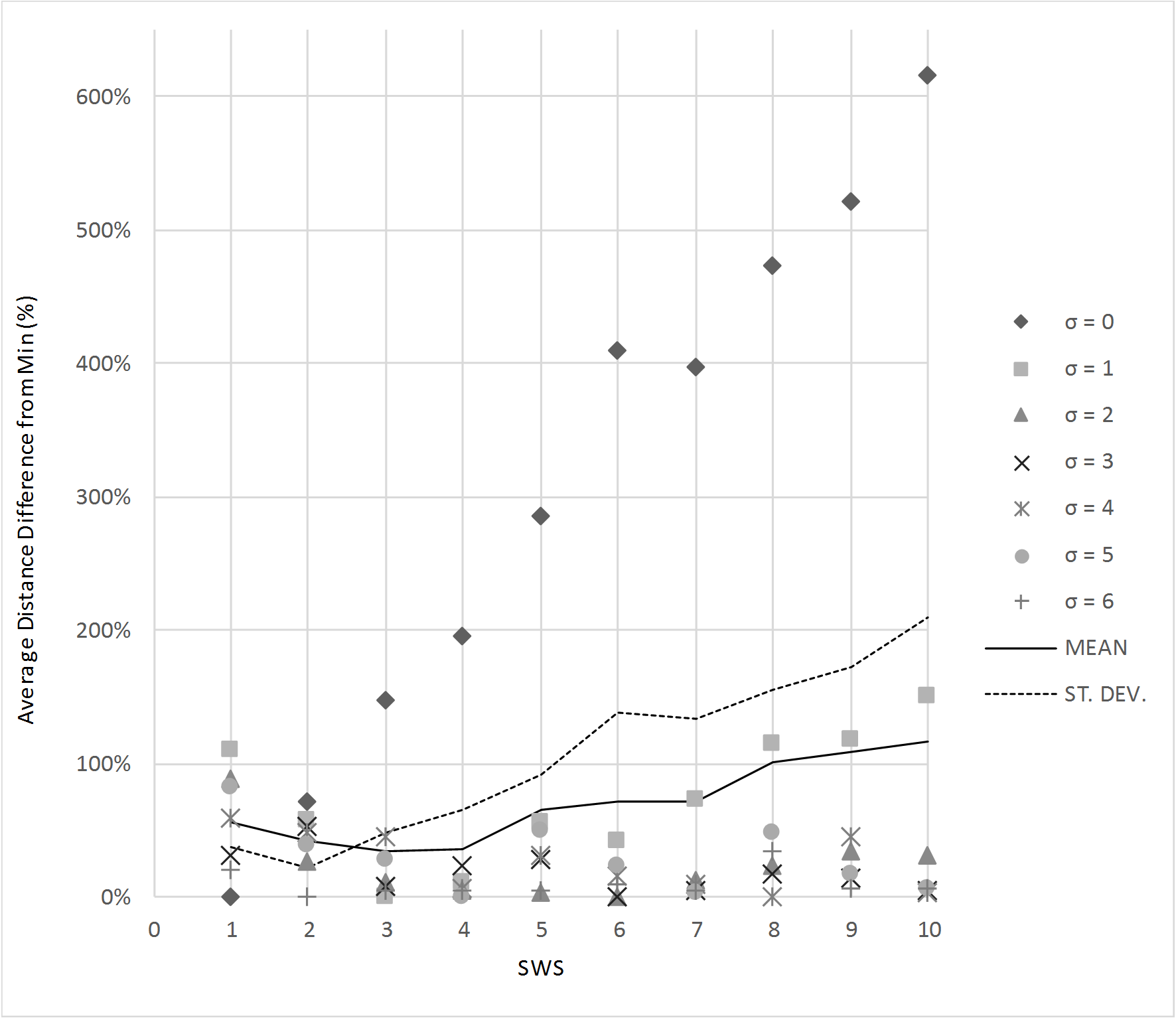}
 \caption{Difference from the minimum average distance between robot and target versus SWS values of the Hot-Cold algorithm, for different standard deviation values of noise due to fading ($\sigma$).}
 \label{fig:diff-average_distance_v3}
\end{figure}

\begin{figure}
 \centering
 \includegraphics[width=\textwidth]{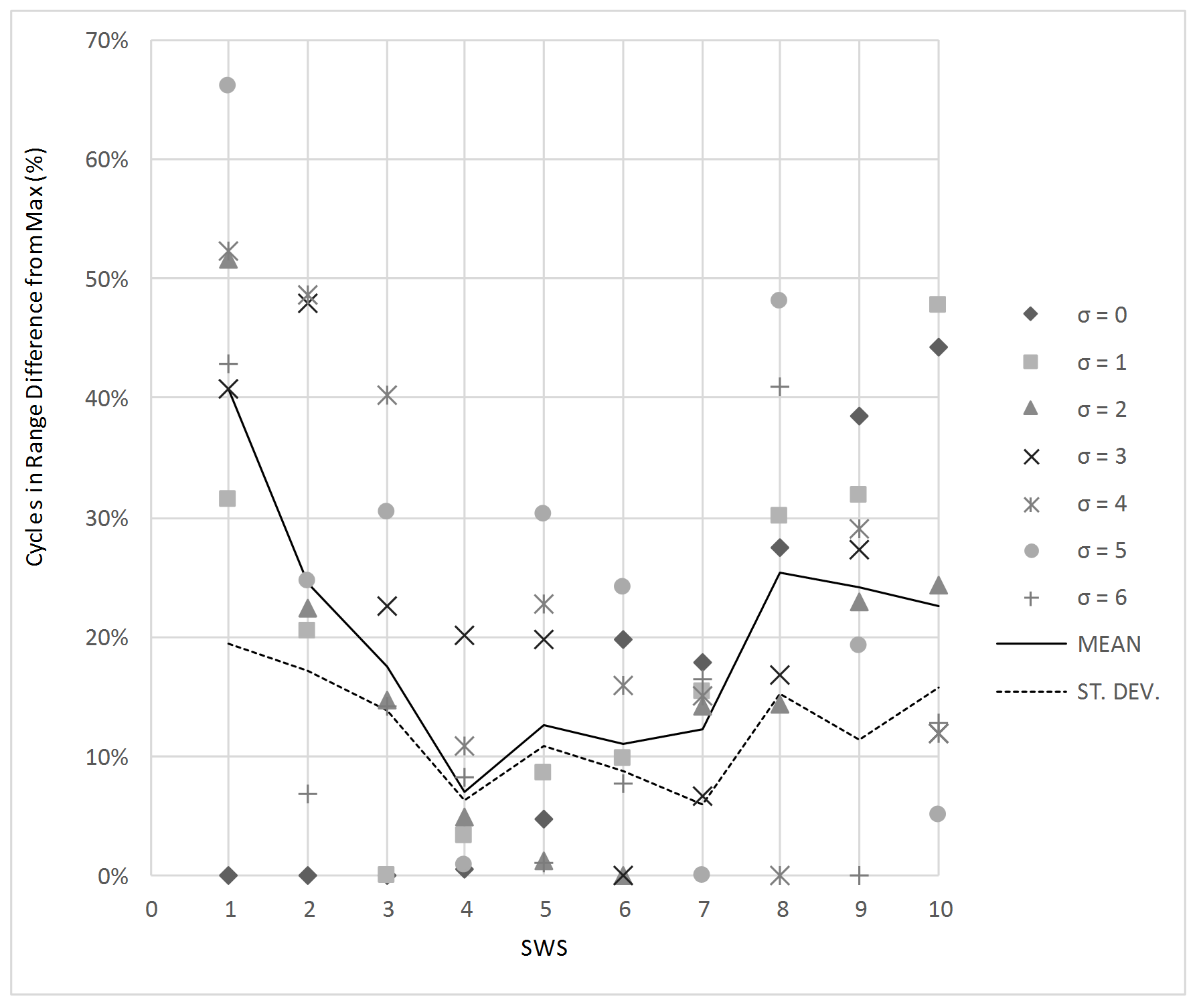}
 \caption{Difference from the maximum cycles the robot stays in communication range with the target versus SWS values of the Hot-Cold algorithm, for different standard deviation values of noise due to fading ($\sigma$).}
 \label{fig:diff-cycles_in_range}
\end{figure}

\begin{figure}
 \centering
 \includegraphics[width=\textwidth]{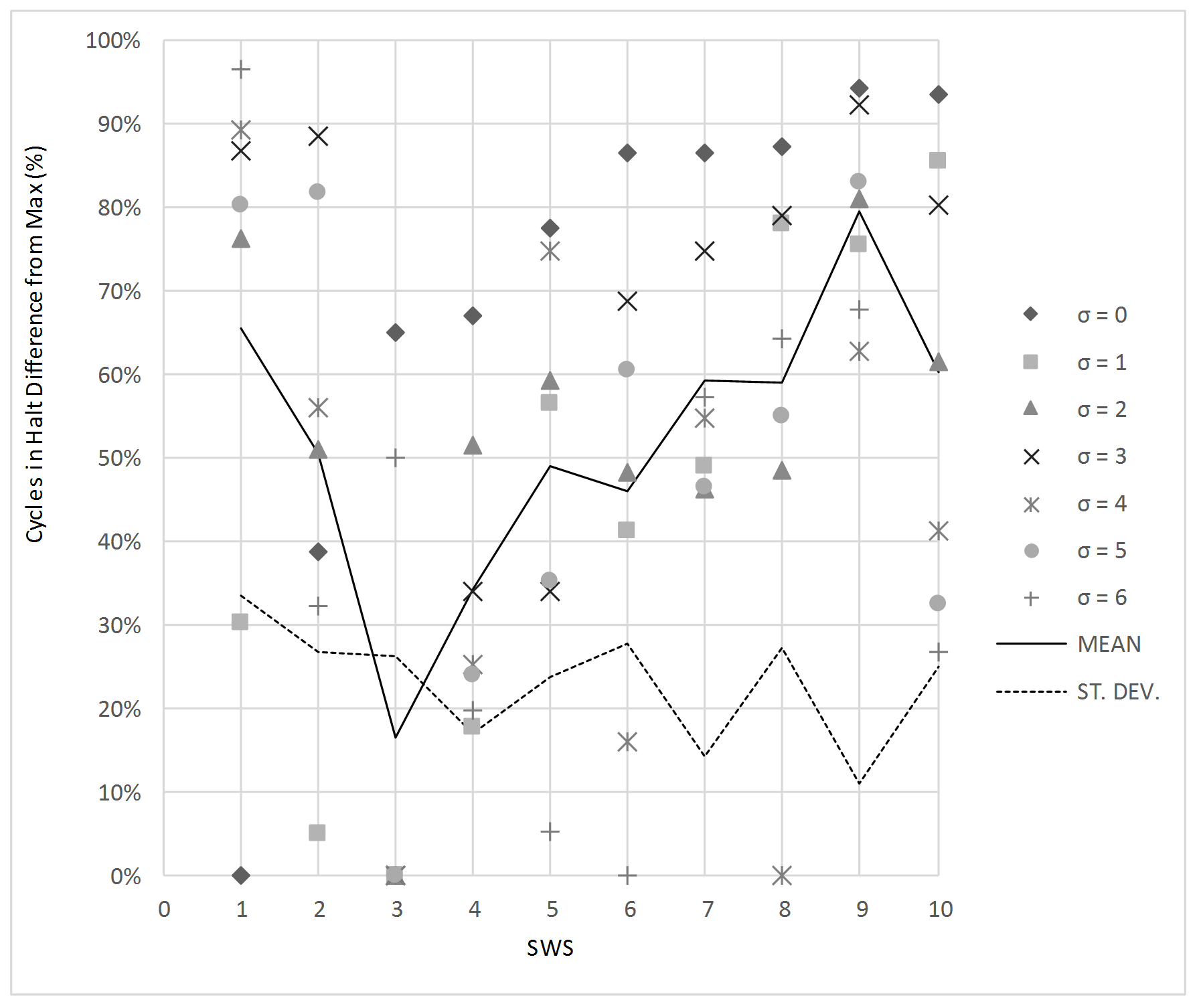}
 \caption{Difference from the maximum cycles the robot stays in halt distance from the target versus SWS values of the Hot-Cold algorithm, for different standard deviation values of noise due to fading ($\sigma$).}
 \label{fig:diff-cycles_in_halt}
\end{figure}

In our effort to identify the optimal SWS values for the Hot-Cold algorithm, we have initially run simulations for SWS ranging from 1 to 10 and for standard deviation of noise due to fading ($\sigma$) ranging from 0 to 6. For each SWS value, the minimum Average Distance is identified, as well as the Average Distance corresponding to different $\sigma$ values. In Figure  \ref{fig:diff-average_distance_v3}, we plot the difference of each Average Distance value from the minimum value, along with the mean and standard deviation. It can be seen that on average the algorithm achieves smallest differences from the minimum distances for SWS values in the range of 3 to 7. In the place of "Average Distance", Figure  \ref{fig:diff-cycles_in_range} depicts "Cycles in Range" and Figure  \ref{fig:diff-cycles_in_halt} depicts "Cycles in Halt", while considering the difference from the maximum value. The former shows that the robot stays more time in range for SWS values lying in the range of 4 to 7. Similarly, Figure  \ref{fig:diff-cycles_in_halt} reveals that the robot reaches halt distance more times with SWS values ranging from 3 to 6.

\begin{figure}
 \centering
 \includegraphics[width=\textwidth]{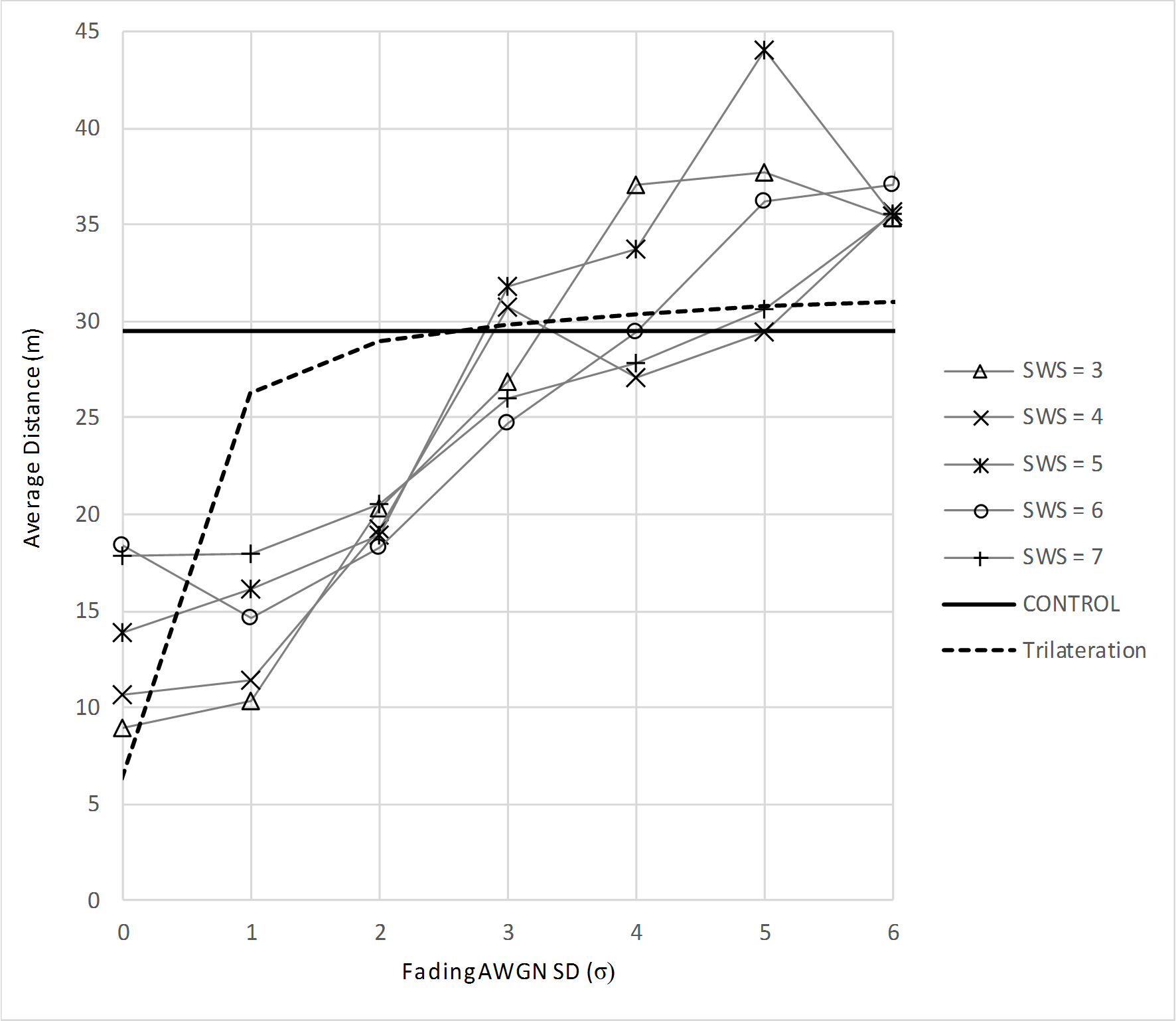}
 \caption{Average distance between robot and target versus the standard deviation of noise due to fading ($\sigma$), for different SWS values of the Hot-Cold algorithm, the Trilateration algorithm, and the control case.}
 \label{fig:average_distance}
\end{figure}

The first part of the simulation-based evaluation of the introduced Hot-Cold algorithm has shown that on average highest following efficiency is achieved for SWS values in the range of 3 to 7. Hence, these are the SWS values that we are using to compare Hot-Cold versus a reference target following algorithm. The simulation results about the "Average Distance" metric are presented in Figure  \ref{fig:average_distance}. The chart plots "Average Distance" as a function of the standard deviation ($\sigma$) of the Additive White Gaussian Noise (AWGN) used for modeling signal fading for different SWS values of the Hot-Cold algorithm as well as the Trilateration target following algorithm and a control case. The latter refers to the case that the robot is completely static, staying in its original position throughout the whole duration of the simulation, which results in average distance of 30 meters. As expected, lower noise yields shorter following distance. Moreover, the results reveal that for increased noise with $\sigma$ equal to 5 or higher, RSSI-based target following algorithms become too inaccurate and they perform even worse than the control case. It should be noted though that this behavior is also due to the fact that while the target is constrained within the simulation space limits, the robot is free to move even beyond those limits, which causes large distances in case of too inaccurate following. However, for lower $\sigma$ values, Hot-Cold outperforms Trilateration, except from the unrealistic case of noise-free signal ($\sigma$ = 0).

\begin{figure}
 \centering
 \includegraphics[width=\textwidth]{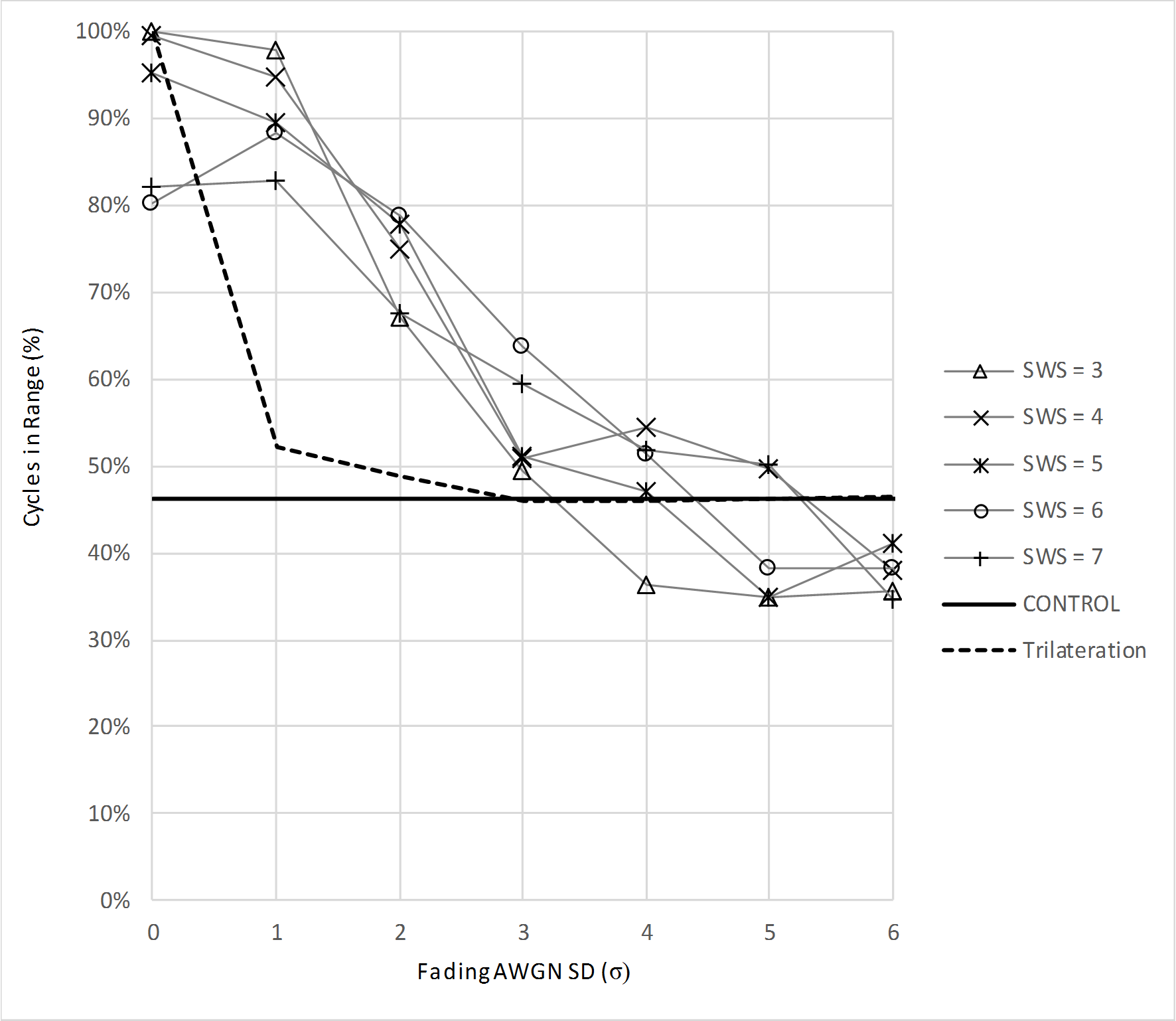}
 \caption{Simulation cycles during which the robot stays within the target's communication range versus the standard deviation of noise due to fading ($\sigma$), for different SWS values of the Hot-Cold algorithm, the Trilateration algorithm, and the control case.}
 \label{fig:cycles_in_range}
\end{figure}

\begin{figure}
 \centering
 \includegraphics[width=\textwidth]{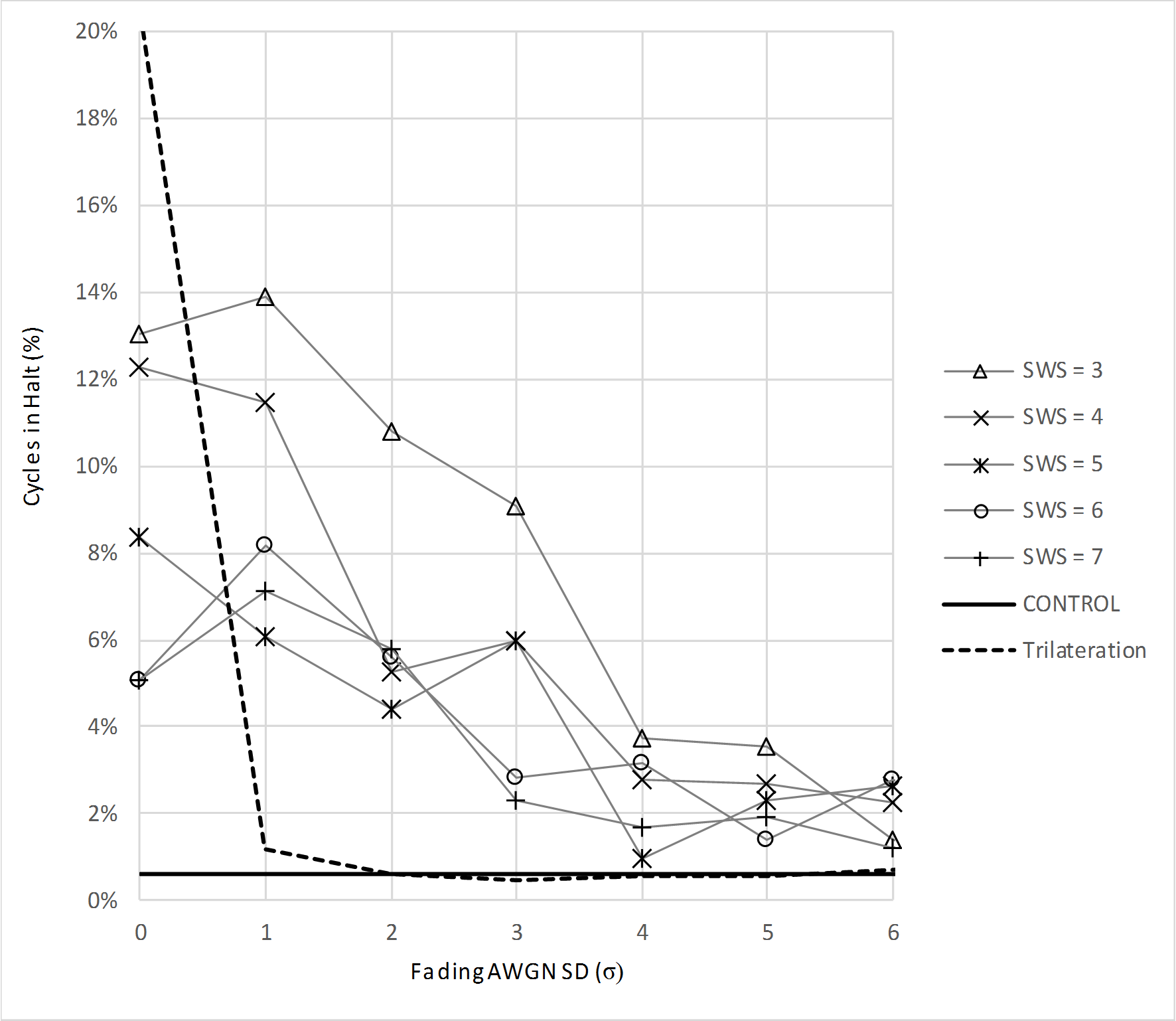}
 \caption{Simulation cycles during which the robot stays within halt distance from the target versus the standard deviation of noise due to fading ($\sigma$), for different SWS values of the Hot-Cold algorithm, the Trilateration algorithm, and the control case.}
 \label{fig:cycles_in_halt}
\end{figure}

Similar conclusions can be drawn when comparing "Cycles in Range". Figure  \ref{fig:cycles_in_range} shows that the robot can stay longer in communication range when Hot-Cold is used for noise levels lower than six standard deviations. For higher levels of noise, staying still (control case) would actually perform better than trying to follow. In the unrealistic case of noise absence, both Hot-Cold and Trilateration can achieve 100\% simulation cycles in range. Reaching halt distance from target (i.e. within 3 meters for our simulations) is even more challenging for the target following process. It can be seen in Figure \ref{fig:cycles_in_halt} that the Trilateration curve overlaps with the control curve at almost 0\% for fading due to shadowing with $\sigma$ higher than 1. In fact, excluding the unrealistic case of $\sigma$ equal to 0, Hot-Cold manages to drive the robot to halt distance clearly more frequently than Trilateration.

\begin{figure}
 \centering
 \includegraphics[width=\textwidth]{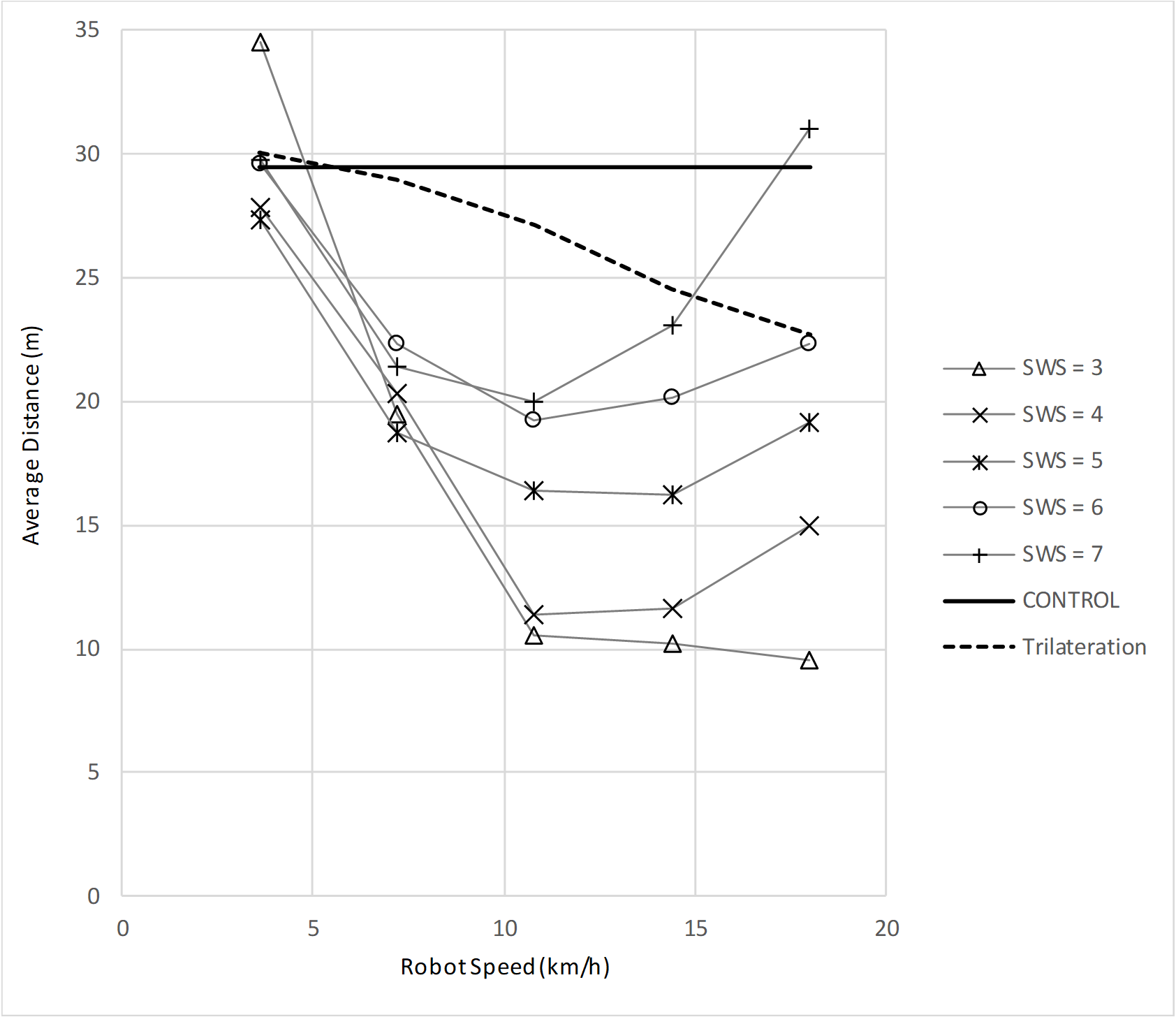}
 \caption{Average distance between robot and target versus the robot speed, for different SWS values of the Hot-Cold algorithm, the Trilateration algorithm, and the control case (standard deviation of noise due to fading ($\sigma$) is set to 2).}
 \label{fig:distance-speed}
\end{figure}

As last part of the simulation-based evaluation of the target following scheme, we investigate the impact of the robot-target relative speed on the tracking efficiency. Figure  \ref{fig:distance-speed} presents the average distance between the target and the robot for different values of robot speed, keeping target’s speed fixed at 3.6 km/h and signal fading $\sigma$ equal to 2. The minimum considered robot speed is 3.6 km/h, since following a target which moves slower than it does not really make sense. The respective control and trilateration values are also plotted as references. It is evident that there is a relation between the robot speed and the SWS value. This is attributed to the distance covered by the robot relatively to the target and the spatial frequency of the RSSI readings. In more detail, since RSSI readings occur at fixed time periods of 0.5 sec, traveling at higher speeds leads to longer distances covered between the readings. In that manner, Hot-Cold decisions become more accurate, thus, tracking becomes more efficient (lower average robot-target distance). However, if the robot moves too fast, then it travels too far before making a rotation decision, which degrades tracking efficiency. This behavior is intensified by the impact of the SWS value, since smaller sample windows lead to more frequent but less accurate decisions. The results reveal that for each value of robot speed there is an optimal SWS which minimizes the average distance. For instance, in the context of the considered values, when the robot moves at 18 km/h, the optimal size of the samples window is 3, assuming that the target speed is 3.6 km/h and the signal fading $\sigma$ is 2.

The results of the conducted simulations for the evaluation of the proposed Hot-Cold algorithm have provided insights for parameters' optimization and conclusions on performance through a comparative study. The Trilateration algorithm is employed as reference and is shown to perform excellent in ideal condition, when there is actually no fading due to shadowing. In all other cases, it fails to drive the robot close to the target as efficiently as Hot-Cold does. The reason is that Trilateration is based on accurate calculation of the robot-target distance according to the received signal strength. This approach provides perfect results in the unrealistic case of a noise free channel, but degrades fast as fading increases. On the other hand, the introduced algorithm provides target following capabilities based on relative signal differences after storing values in the Samples Window. As a result, Hot-Cold is more tolerant to random fading effects, managing to effectively turn the robot towards the moving target. Of course, when signal gets too unstable with great deviations ($\sigma$ higher than 5), RSSI-based target following becomes too unreliable and quite infeasible. Most efficient following is shown to be possible for SWS values between 3 and 7. 

\subsection{Experimental Testbed}
In order to perform real-world experiments on the introduced scheme, we have developed a testbed which enables following an individual and providing remote real-time access to her e-health data in the form of a service. 

\subsubsection{Robotic Device Software}
%(del.6: 4.1 with focus on Figure 36)
In this subsection, we focus on the robotic device software, which includes the Hot-Cold algorithm. For this testbed, the robot software \cite{tlagkas_tlagkashot-cold_prototype_2020} was developed in the Eclipse Integrated Development Environment using Java and the LeJOS framework \cite{noauthor_lejos_nodate}. According to the system architecture, the program is executed in a Raspberry Pi that constitutes the sensor node and is mounted on the robotic device. It controls the robot over a USB connection with the robot processing unit. 
%The developed software is object-oriented. A behavior programming approach was followed to control the robot.

% Since RSSI-based target following is greatly affected by signal power variations due to the environment or other physical phenomena, we have enhanced the robot's target following ability with infrared-based tracking. The main idea is that as long as the robotic device has Line Of Sight (LOS) with the infrared emitter attached to the followed target, it uses the corresponding sensor to accurately track it. When it gets out of LOS or moves away from the short infrared range, the RF-based target following scheme (Hot-Cold) kicks in to retain radio connectivity. 
Moreover, for practical reasons we have implemented in the robot an obstacle avoidance mechanism, which relies on two ultrasonic sensors positioned at the two corners (left, right) of the robot's lower front side that can detect obstacles at a distance up to 255 cm at an angular range of approximately $\pm90^{\circ}$. The concept of the respective developed algorithm is twofold: a) avoid moving towards the same obstacle repetitively, b) avoid following a direction which is almost opposite to the one already followed. For instance, if the robot meets a wall at an angle, it should always avoid it retaining the same direction and not turning back. The readings of the ultrasonic sensors are constantly checked. The corresponding steps are: 
\begin{itemize}
    \item 
    If the readings of both sensors get lower than 25 cm, then the robot travels backwards by 10 cm and then rotates by $45^{\circ}$.
    \item
    Else if the reading of just the right sensor gets lower than 25 cm, then the robot travels backwards by 10 cm and then rotates by $+10^{\circ}$ (i.e. left direction).
    \item
    Else if the reading of just the left sensor gets lower than 25 cm, then the robot travels backwards by 10 cm and then rotates by $-10^{\circ}$ (i.e. right direction).
\end{itemize}
\noindent It is noticed that the specific obstacle avoidance scheme allows the robot to navigate both in wider as well as in narrower spaces. It never requires backwards movement for more than 10 cm, while it can drive the robot through a corridor as narrow as 70 cm (so that it can move in straight line between the walls without triggering the sensors). Of course, it should be clarified that the obstacle avoidance algorithm is fully configurable and replaceable, since it is not part of the main focus of this work. In fact, it is considered as part of the lower layer motion control, whereas the proposed Hot-Cold tracking algorithm operates on top of it indicating the general direction towards the target.

%The UML diagram of the Java classes the robot program includes is presented in Figure \ref{fig:UML_diagram}. It is noted that the RotateRandomly behavior in fact realizes rotation at 137 degrees according to the Hot-Cold algorithm, which is tuned at SWS value of 4. 

%\begin{figure}
% \centering
% \includegraphics[width=0.9\textwidth]{UML_diagram_v3.png}
% \caption{UML diagram of the software developed in the prototype target following robotic %device, realizing the Hot-Cold algorithm.}
% \label{fig:UML_diagram}
%\end{figure}

\subsubsection{Testbed Setup}
%(del.6: 4.3)
With the completion of the testbed, three experimental scenarios were set up, each one under two different radio propagation conditions. The tested case was monitoring in real-time the vital sings of a followed individual, both locally and remotely within the context of an IoT architecture. The role and properties of each entity of the testbed are described below:
\begin{itemize}
\item \textit{Prototyped Sensor Platform}: The sensor platform was implemented and tested using an Arduino Uno board in combination with an e-health kit including a number of e-health sensors for estimating airflow, temperature, skin conductance, skin resistance, heart rate, SPO2, electrocardiogram, and body position. We equipped the board with a wireless network module featuring IEEE 802.15.4 \cite{noauthor_ieee_2016} for broadcasting sensor data.
\item \textit{Prototyped Tracking Device}: The device following the monitored individual is a mobile robotic device, equipped with an ultrasound sensor for obstacle avoidance.
% and an infrared sensor for short-range tracking. 
\item \textit{Prototyped Sensor Node}: A Raspberry Pi 3 realizes the sensor node. It is equipped with an IEEE 802.15.4 module (through a conversion bridge) to communicate with the sensor platform. Moreover, it uses an IEEE 802.11 USB dongle to provide access to sensor data through relaying. Each sensor node performs MANET routing using the "Better Approach To Mobile Adhoc Networking" protocol \cite{johnson_simple_2008} and is attached on the robotic device.
%\item \textit{Prototyped Relay Node}: The relay node is implemented %through a Raspberry Pi 3 which includes two IEEE 802.11 wireless USB %dongles to receive sensor data from sensor nodes and relay them to the %gateway. It forms a MANET along with the sensor node, using the "Better %Approach To Mobile Adhoc Networking" as routing protocol %[\textbf{reference for BATMAN}]. 
\item \textit{Prototyped Biospace/Biobot}: The EDBO Biospace was developed via JADEX as a Java program running in the relay node. The Biospace creates a Biobot providing the corresponding sensor data service, which is made discoverable to the whole Global Network of the IoT architecture through the provided service registry.
%\item \textit{Prototyped External System}: The external system was %implemented in Java and executed in the sensor node.
\item \textit{Prototyped End User System}: The end user system was realized as a Java-based client software running in a tablet, which is able to discover the created Biobots and access their sensor data service.
\end{itemize}

The testbed was deployed in a closed-space sports University facility of dimensions 35m x 40m (totaling 1400 m$^2$), where the Cartesian axes origin (0, 0) is placed at the top left corner. A robotic device equipped with a sensor node moves at 10 km/h and performs Hot-Cold tracking (with parameters $SWS=4$ and $\phi=137^{\circ}$) of a person carrying a sensor platform, who walks at 5 km/h when moving. The edge-to-edge front wheel distance (i.e. the track width) of the robotic device is 17.5 cm and the duration of a direction change (stop-rotate-start) is on average 2.5 sec. The sensor data are forwarded over an IEEE 802.11 MANET through a Raspberry Pi relay node to a laptop that constitutes the end user system. Four iterations of 60 sec are performed for each experimental scenario, which are set up as follows:
\begin{itemize}
    \item \textit{1st Scenario}: This is considered as control scenario, where the target remains static at position (5, 5). The robot's starting position is (30, 35) facing away from the target (direction at $50^{\circ}$).
    \item \textit{2nd Scenario}: The target moves in straight line starting at point (5, 5) and in 28 sec reaches the destination point (50, 35). The robot's starting position is (30, 5) with initial direction $0^{\circ}$.
    \item \textit{3rd Scenario}: The target moves in zigzag starting at point (5, 5), visiting  after 10 sec the first waypoint (5, 11.5), 20 sec later it reaches the second waypoint (30, 25), and 20.5 sec later it reaches the destination (5, 35). The robot's starting position is (30, 5) with initial direction $0^{\circ}$.
\end{itemize}

Two experimental evaluations took place for each one of the three scenarios, corresponding to different radio propagation conditions. In detail, the first experiment was performed under normal noise conditions, with no interference intentionally created in the testbed environment. The second experiment performed for each scenario took place under elevated noise conditions using two sources of interference: a pair of Raspberry Pi devices positioned at coordinates (0,0) and (35,40), respectively. These two devices were equipped with a 5 dBi omni-directional antenna and were configured to exchange data over an IEEE 802.11n link operating at channel 6, which has a central frequency of 2437 MHz and bandwidth of 20 MHz, and transmit at 13 dBm TX power. It is noted that the specific communication directly interferes with the IEEE 802.15.4 link connecting the target to the robotic device, which is set at frequency 2435 MHz that corresponds to channel 17 of the specific standard.

\subsubsection{Testing Results}
\begin{figure}
 \centering
 \includegraphics[width=0.9\textwidth]{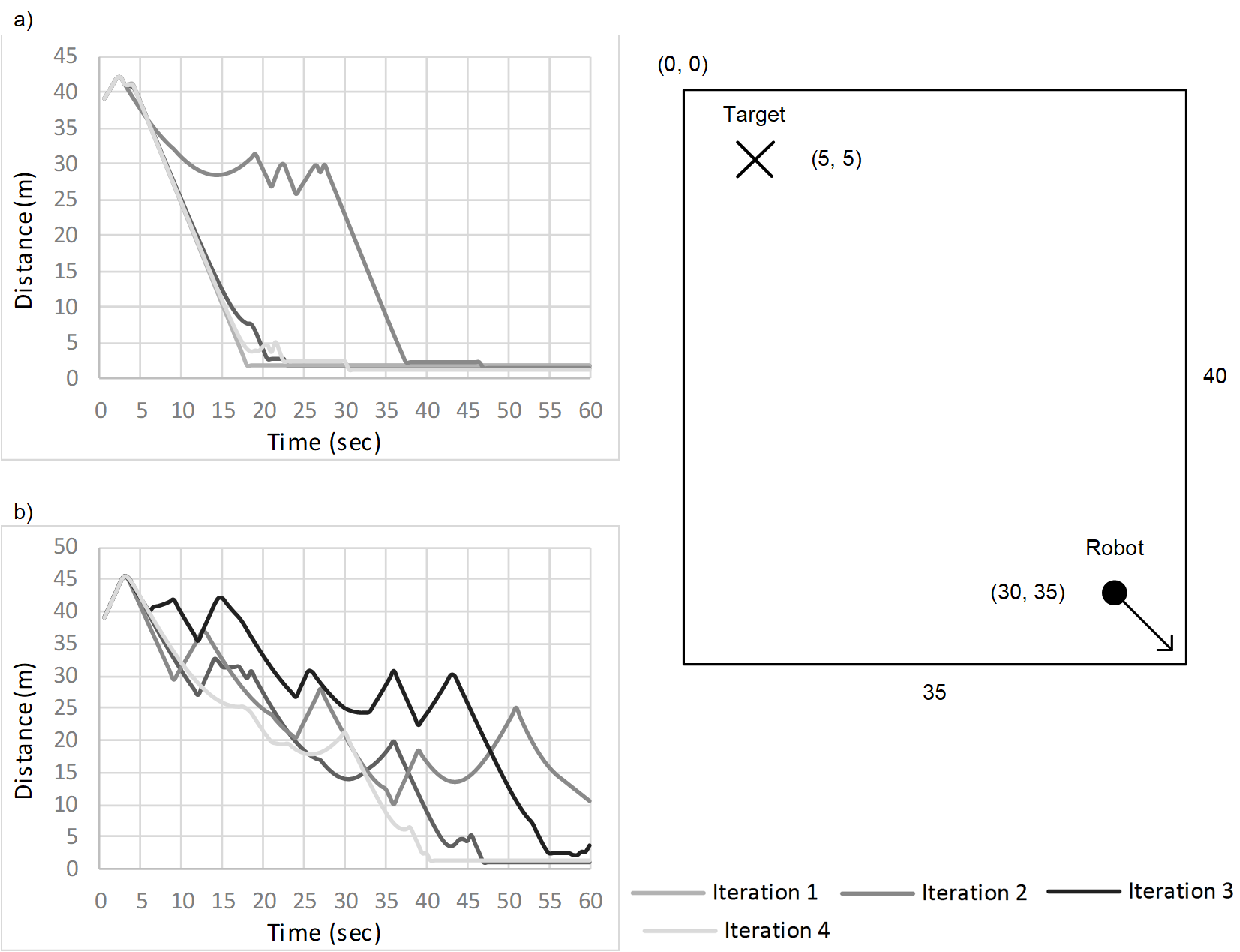}
 \caption{Distance between target and robot adopting Hot-Cold 
tracking versus time in the 1st scenario, under a) normal noise and b) elevated noise conditions}
 \label{fig:prey_fixed}
\end{figure}

The experimental results of the robot-target distance for the 1st scenario are presented in Figure \ref{fig:prey_fixed}, along with an illustration of the scenario setup. In Figure \ref{fig:prey_fixed}a, it can be seen that in normal noise conditions, after correcting its direction, the robot manages to reach the static target within the experiment duration. Specifically, in three iterations the target is reached in about 20 sec, while in one iteration the target is reached in about 37 sec. Figure \ref{fig:prey_fixed}b shows that under elevated noise conditions, the robot has to correct its direction several times, due to some erroneous estimates of the signal strength variations caused by the induced interference, however, it still manages to closely approach the target within the duration of the experiment.

\begin{figure}
 \centering
 \includegraphics[width=0.9\textwidth]{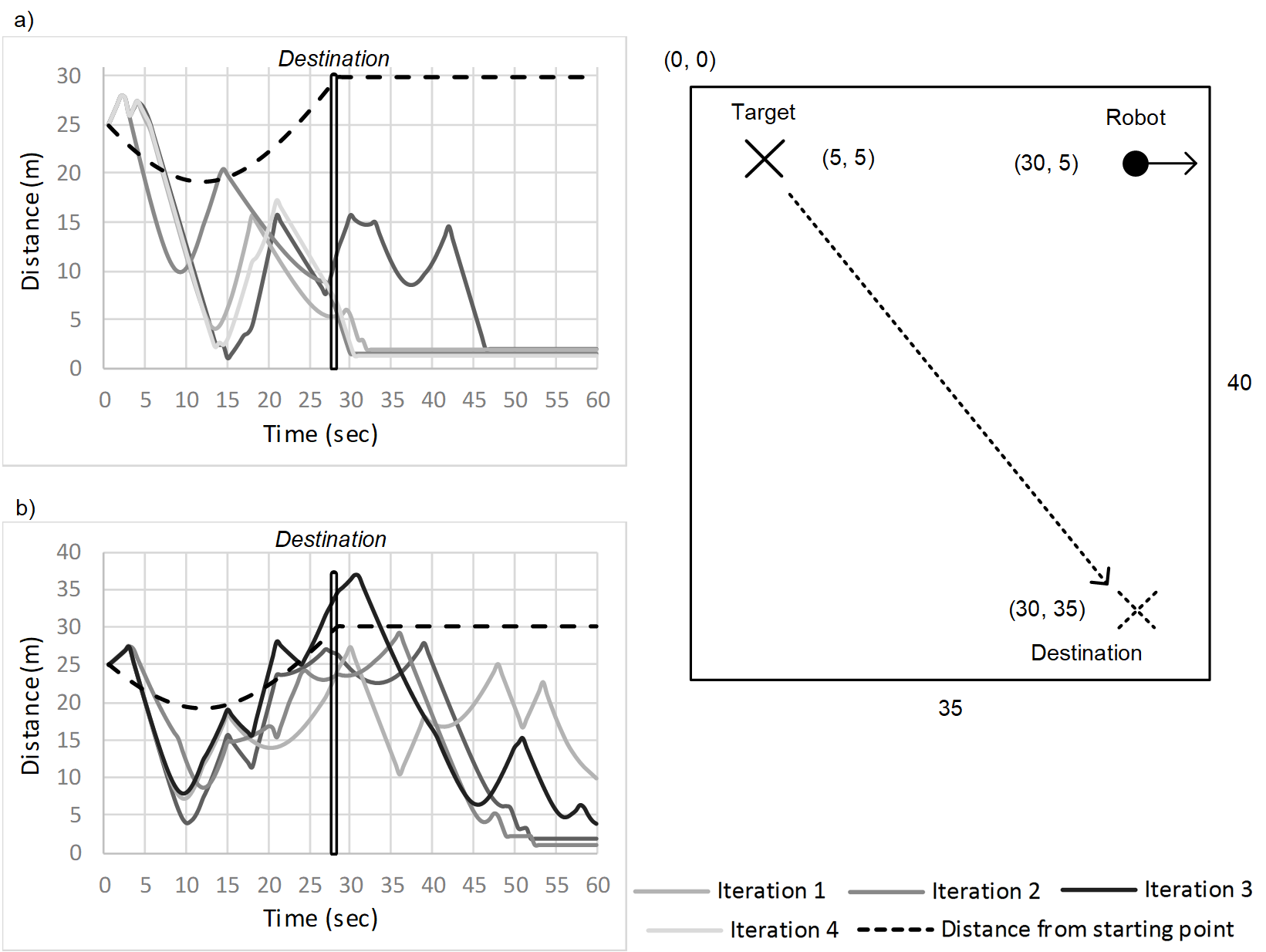}
 \caption{Distance between target and robot adopting Hot-Cold 
tracking versus time in the 2nd scenario, under a) normal noise and b) elevated noise conditions}
 \label{fig:prey_straight}
\end{figure}

In the 2nd scenario, the charts in Figure \ref{fig:prey_straight} also present target's distance from the robot's starting point (dashed line). It is evident that under normal noise conditions (Figure \ref{fig:prey_straight}a) the robot initially corrects its direction and reaches the target about 2 sec after the latter arrives at destination (at time 28 sec), except from one iteration that required sixteen more seconds. Under elevated noise conditions, it can be seen that the robot has to realign its trajectory several times, but eventually manages to stay in the target's proximity, as show in Figure \ref{fig:prey_straight}b.

\begin{figure}
 \centering
 \includegraphics[width=0.9\textwidth]{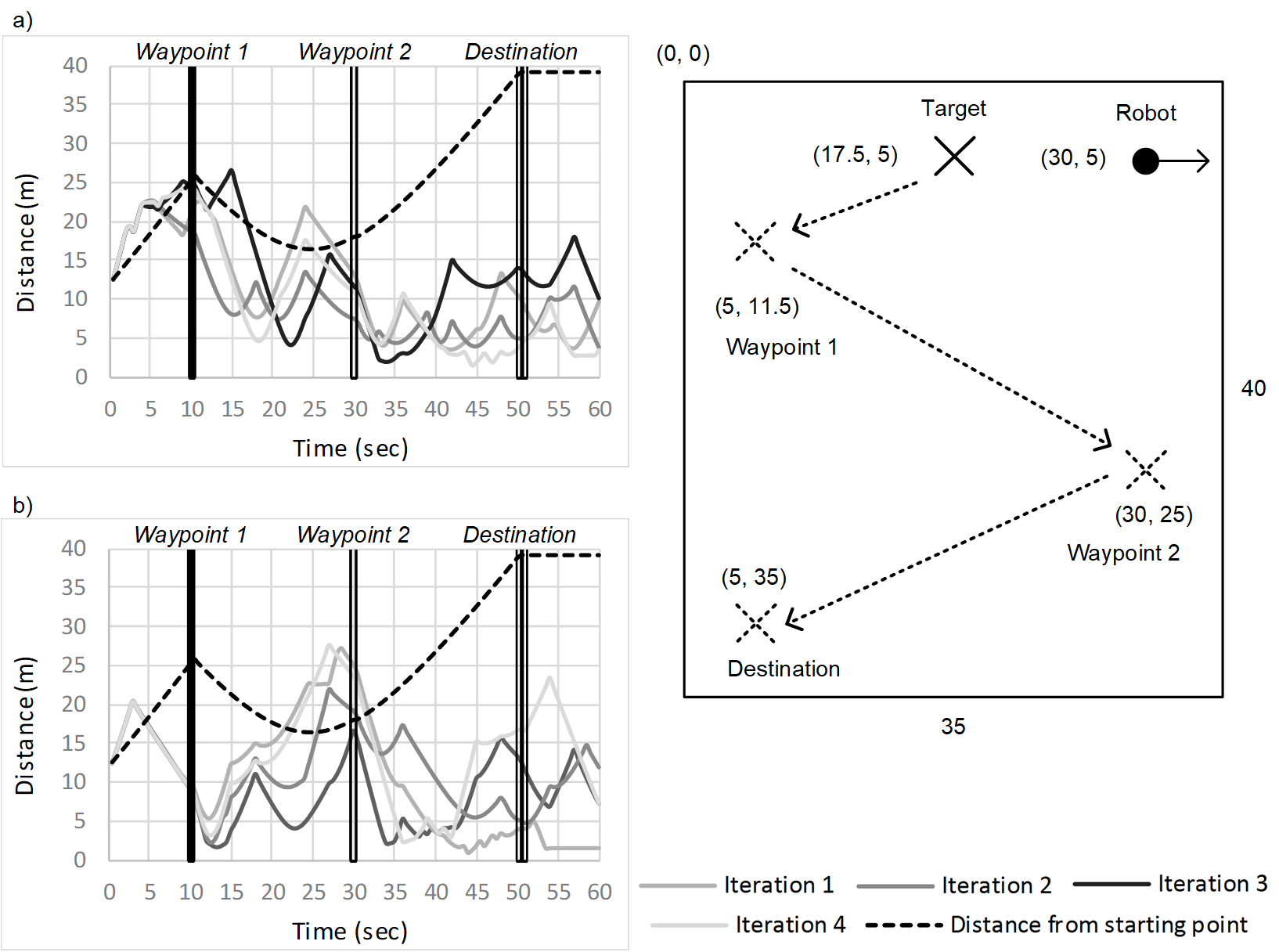}
 \caption{Distance between target and robot adopting Hot-Cold 
tracking versus time in the 3rd scenario, under a) normal noise and b) elevated noise conditions}
 \label{fig:prey_zigzag}
\end{figure}

Probably the most challenging scenario is the third one, with the corresponding results for normal noise conditions presented in Figure \ref{fig:prey_zigzag}a. Until the first waypoint, the target moves almost opposite from the robot's initial direction, causing temporal distance increment. After that, the robot approaches, reaching minimum distance right after the target's arrival at the second waypoint. Then, the robot manages in all iterations to maintain proximity, however, the 9.5 sec that the target remains static at the destination point is not sufficient time for the robot to stay stably close to the target. A similar behaviour is also observed under elevated noise conditions, as presented in Figure \ref{fig:prey_zigzag}b. It is evident that in the presence of excessive interference, it is quite difficult for the robot to stay close to the target when the latter reaches the second waypoint, however, it eventually manages to approach closely by the end of the experiment.

Conclusively, in all three scenarios, the Hot-Cold algorithm has succeeded in its goal of adjusting robot's trajectory to keep approaching the target, affected of course by the mobility pattern. Artificially increasing the radio interference led to degradation of the tracking performance (longer intervals before reaching the target), however, the robot managed to stay in target's close proximity. During the experiments, end-to-end connectivity to the sensor platform was maintained, with all sensor data collected every 0.5 sec successfully relayed to the end user system.

In general, the tracking environment may affect the behavior of the Hot-Cold scheme in various ways. The structure of the considered scene has a direct impact on the RF signal propagation as well as on the target’s mobility pattern and the robot’s following abilities. Specifically, the presence of multiple obstacles (such as in an urban environment) can create high shadowing and multipath fading effects which lead to less reliable RSSI estimates, hence, to less accurate rotation decisions. On the other hand, a relatively open scene ensures weaker fading, thus, more efficient tracking decisions based on the received signal strength fluctuations. Moreover, heavily obstructed paths make it difficult to maintain a consistent tracking course and put most of the pressure on the adopted obstacle avoidance technique, which however is not an internal component of the introduced Hot-Cold algorithm, but a complementary one. Generally, as expected, all elements that may affect signal propagation (such as objects’ reflection and refraction factors) and/or mobility have an impact on the efficiency of the introduced technique. The presented evaluation results have revealed that in environments where the tracking robot can effectively move via obstacle avoidance and the standard deviation of the experienced Gaussian fading is lower than 5, the proposed scheme can ensure effective tracking performance.

\section{Conclusions and Future Directions}
% CONTENT COMMENTS:
% Including scaling up to n patients / m robots
In this paper, we have primarily introduced a new algorithm for following mobile monitored targets/individuals in the context of an IoT system. The devised technique, called Hot-Cold, is able to ensure proximity maintenance by the tracking robotic device solely based on the strength of the RF signal broadcasted by the target to communicate its sensors' data. Possible applications of such a tracking technique are quite promising and include the sustainment of communication links for monitoring purposes in dynamic environments with limited or unavailable network infrastructure.
%, such as in the case of disaster scenarios. 
The monitoring information is made available over a bio-inspired IoT architecture, which allows flexible creation and discovery of sensor-based services. 

For the identification of the optimal rotation angle employed by the tracking robot, a complete analysis was conducted in four steps: geometrical analysis, numerical analysis, exhaustive-simulation analysis, and convergence analysis. The analytical results reveal that performance optimization is achieved for Hot-Cold at a rotation angle of $\sim$137 degrees. An in-depth evaluation of the proposed technique was performed through simulations and in comparison with the well-known concept of trilateration-based tracking. The simulation results have identified the optimal configuration for Hot-Cold key parameters and have shown that it achieves superior performance for realistic levels of signal fading due to shadowing. The evaluation part is completed with the presentation of a testbed, which demonstrates the proposed IoT system concept. All key components are thoroughly described, focusing on the target following aspect. The conducted experiments show the operability of the overall approach and especially focus on the effectiveness of the tracking technique.

Future work involves the optimization of the tracking technique for generalized target following scenarios in the context of IoT. For instance, we intend to investigate combinations of different numbers of tracking devices following one or more monitored targets in a cooperative manner. Moreover, possible applications can be extended from mobile tracking robots to Unmanned Autonomous Vehicles (UAVs - drones). In general, the related potential extensions in terms of candidate applications and functionality enhancements are numerous and very promising for the future of IoT and they definitely worth further exploration.

%\section*{References}

\bibliography{Following_Mobile_IoT_Elsevier_PMC_R1.bib}

\end{document}